%% file: manuscript.tex
\documentclass[3p, sort&compress]{elsarticle}        

\usepackage{float}

\journal{Computer Physics Communications}

\usepackage{amsmath, amssymb, subfig}

\usepackage{hyperref}
\hypersetup{unicode}

\def\defeq{\mathrel{\mathop:}=}
\def\eqdef{=\mathrel{\mathop:}}
\newcommand{\gl}[2][]{\begin{equation#1} #2 \end{equation#1}}

\DeclareMathOperator{\floor}{floor}

\begin{document}               
\begin{frontmatter}

\title{%
  EZ: An Efficient, Charge Conserving Current Deposition Algorithm for
  Electromagnetic Particle-In-Cell Simulations
}  

\author[1,2]{Klaus Steiniger\corref{mycorrespondingauthor}%
  \fnref{equalcontribution}}
\fntext[equalcontribution]{These authors contributed equally}
\cortext[mycorrespondingauthor]{Corresponding author}
\ead{k.steiniger@hzdr.de}

\author[1]{Rene Widera\fnref{equalcontribution}}

\author[1]{Sergei Bastrakov}
\author[1,2]{Michael Bussmann}
\author[3]{Sunita Chandrasekaran}
\author[4]{Benjamin Hernandez}
\author[3]{Kristina Holsapple}
\author[5]{Axel Huebl}
\author[1]{Guido Juckeland}
\author[1]{Jeffrey Kelling}
\author[3]{Matt Leinhauser}
\author[1]{Richard Pausch}
\author[4]{David Rogers}
\author[1,6]{Ulrich Schramm}
\author[7]{Jeff Young}
\author[1]{Alexander Debus}

\address[1]{Helmholtz-Zentrum Dresden -- Rossendorf, Bautzner Landstraße 400, 01328 Dresden, Germany}
\address[2]{CASUS -- Center for Advanced Systems Understanding, Untermarkt 20, 02826 G\"{o}rlitz, Germany}
\address[3]{Department of CIS, University of Delaware, Newark, DE 19716, USA}
\address[4]{National Center for Computational Sciences, Oak Ridge National Laboratory, Oak Ridge, TN 37831, USA}
\address[5]{Lawrence Berkeley National Laboratory, Berkeley, CA 94720, USA}
\address[6]{Technische Universit\"{a}t Dresden, 01062 Dresden, Germany}
\address[7]{Georgia Tech, School of Computer Science, Atlanta, GA 30332, USA}

\begin{abstract}
We present EZ, a novel current deposition algorithm for particle-in-cell (PIC) simulations. EZ calculates the current density on the electromagnetic grid due to macro-particle motion within a time step by solving the continuity equation of electrodynamics.
Being a charge conserving hybridization of \textbf{E}sirkepov's method and \textbf{Z}igZag, we refer to it as ``EZ'' as shorthand for ``Esirkepov meets ZigZag''.
Simulations of a warm, relativistic plasma with PIConGPU show that EZ achieves the same level of charge conservation as the commonly used method by Esirkepov, yet reaches higher performance for macro-particle assignment-functions up to third-order.
In addition to a detailed description of the functioning of EZ, reasons for the expected and observed performance increase are given, and guidelines for its implementation aiming at highest performance on GPUs are provided.
\end{abstract}


\end{frontmatter}


\section{Introduction}

Electromagnetic particle-in-cell (PIC) ~\cite{hockney1988, birdsall1991} simulation is the most widespread method of modeling kinetic plasma phenomena and is a quasi-gold standard for computer experiments.
PIC simulations have a wide range of applications, such as laser-plasma acceleration~\cite{Fonseca2008,Burau2010,Lehe2015a,Couperus2017,Obst2018,Derouillat2018,Debus2017a,Raj2020,Kurz2020,CouperusCabadag2021}, laser-driven light sources~\cite{Haugbolle2013,Pausch2014a}, inertial confinement fusion~\cite{Pukhov1999}, and astrophysical plasma phenomena~\cite{Nishikawa2009,Bussmann2013,Grismayer2013a,PauschPRE2017}.
Many state-of-the-art PIC codes have memory requirements that need high performance computing resources.
Notably, simulations of solid-density plasmas commonly require substantial amounts of computing and storage capacity of current top supercomputers~\cite{Hilz2018,Myers2021}.

Most of the computational time of a PIC iteration is spent on the \emph{current deposition} (CD) stage, even in highly optimized codes.
Leinhauser et al. measured that close to 60\,\% of an iteration was spent on the CD stage for a representative plasma physics case simulated by PIConGPU \cite{leinhauser2022}.
The current deposition stage computes the density values on the grid from charged macro-particle motion in the PIC simulation.
This density distribution can determine the new electric and magnetic field strengths in a manner that respects Gauss's law.
Widely adopted charge conserving current depositions schemes were developed by Eastwood \cite{eastwood1991}, Villasenor et al.\ \cite{villasenor1992}, Esirkepov \cite{esirkepov2001}, and Umeda et al.\ \cite{umeda2003}.

The allocations of compute resources for laser-plasma simulations can reach hundreds of million core hours on top supercomputers \cite{praceWWW}.
Considering these high computational demands, an improvement in current deposition run time without loss of precision would significantly reduce resources necessary to carry out these simulations.

Here, we present a novel current deposition method offering compute savings on the scale of such high resource demands.
It is a hybridization of Esirkepov's method~\cite{esirkepov2001} and ZigZag~\cite{umeda2003, yu2013}, referred to as ``EZ'' henceforth.
EZ combines the best of both methods by being exact charge conserving and applicable to arbitrary macro-particle \emph{assignment-functions}\footnote{We follow the naming of \cite{hockney1988}, but use symbol $S$ where they use symbol $W$.
The assignment function is also referred to as \emph{shape factor} by \cite{birdsall1991} and \emph{form-factor} by \cite{esirkepov2001}, both using symbol $S$.
Hierarchy of assignment functions: CIC -- first-order, TSC -- second-order, PQS -- third order.} like Esirkepov's method, and maintaining a reduced computational effort like ZigZag.
EZ can be implemented in existing PIC codes and replaces Esirkepov or ZigZag, resulting in at least similar, and often better, performance.

We implemented and tested EZ to deployment in PIConGPU, a popular PIC code.
Our experiments verify the important charge conservation property of EZ and demonstrate performance improvements over Esirkepov's method across a variety of compute architectures.

PIConGPU is an open source, performance-portable implementation of the fully relativistic 3D3V particle-in-cell method written in C++ \cite{Bussmann2013, picongpuWWW}.
It is used by a steadily growing and world-spanning scientific community, including fields such as particle physics, cancer research, and materials science.
Since PIConGPU builds on the performance portability library \emph{alpaka} \cite{zenker2016, matthes2017}, there is a single source C++ implementation of EZ in PIConGPU, which additionally takes the macro-particle assignment-function as a compile-time template argument.
The template argument allows a generic implementation for all assignment-functions and alpaka enables mapping of the single implementation to all supported compute platforms. This abstraction prevents the need for redundant implementations of the method for different macro-particle assignment-functions and compute platforms.
As a result, EZ remains maintainable and portable, just as PIConGPU itself.

\section{EZ Single Macro-Particle Current Density Calculation Method}
\subsection{Introduction to the Method}
First we introduce the principle idea of the method and why a performance increase is expected in comparison to existing methods.

EZ calculates the total current density in a similar way to existing methods. At each particle-in-cell iteration, EZ obtains the total current density on the grid by adding all single macro-particle contributions on the grid.

EZ's novelty lies in the computation of each individual macro-particle contribution.
In general, the single macro-particle contribution to the current density is defined by the motion of the macro-particle within an iteration step.
At the beginning of current deposition we know the old and new positions of a macro-particle. Existing methods differ in the assumption they make about which path a macro-particle takes from its old to new position.
Whereas Esirkepov's method assumes a straight line motion \cite{esirkepov2001}, ZigZag assumes a non-straight line motion leading over a \emph{relay point} \cite{umeda2003}.
EZ employs a macro-particle path splitting approach similar to ZigZag, but calculates the current density contribution for each segment with Esirkepov's method.
Considering ZigZag has not been generalized to arbitrary macro-particle assignment functions, EZ introduces a novel macro-particle path splitting approach for arbitrary macro-particle assignment functions.

With EZ, the initial and final positions of a macro-particle are used to define a relay point. This relay point is where the path of the macro-particle is split. Then, the macro-particle path is decomposed into two smaller paths: the initial position to the relay point and the relay point to the final position.
EZ then uses Esirkepov's method to calculate the current density of each smaller path.

EZ's path splitting with Esirkepov's method offers improved efficiency compared to Esirkepov's original method in two ways:
\begin{enumerate}
\item Computational loops require fewer iterations to calculate single macro-particle current density components.
\item Macro-particle assignment-functions can be implemented for evaluation only on their support, sparing conditional statements required if care needed to be taken for out-of-support evaluations. Eliminating conditional statements and branching on highly parallel computer hardware avoids significant performance penalties \cite{serialization2023}.
\end{enumerate}

\subsection{Definition of the Relay Point and the Assignment Cell}
\begin{figure}[!tb]
  \centering
  \includegraphics[width=0.45\textwidth]{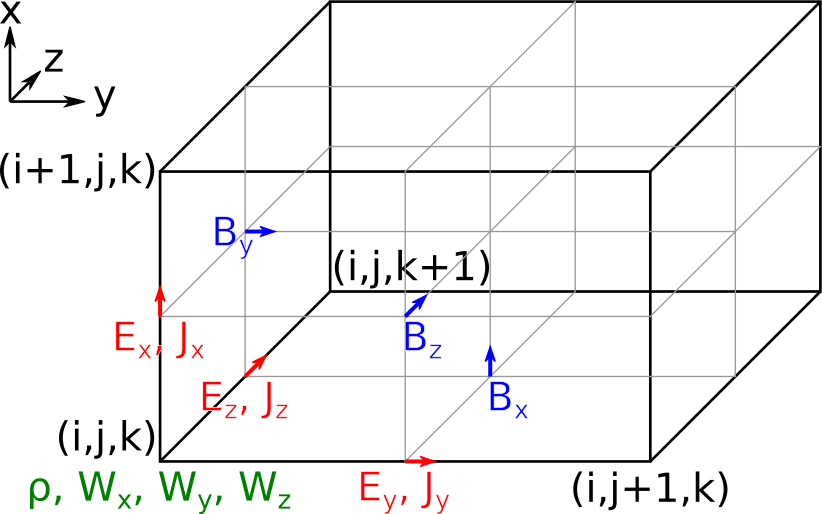}
  \caption{
    Yee cell with charge density $\rho$ and current deposition vector $\mathbf W$ at a cell's origin, components of the magnetic field at the center of faces, components of both electric field and current density halfway along edges.
  }
  \label{fig::YeeCell}
\end{figure}

The trajectory of a macro-particle is split in two parts only if the grid nodes to which its charge is assigned are different before and after movement.
The discrete charge density $\rho \rvert^{n}_{i, j, k}$ (at time step $n$) is calculated on a staggered Yee grid.
The layout of the Yee cell and the location of the charge density values are visualized in fig.\ \ref{fig::YeeCell}.
Charge density values $\rho \rvert^{n}_{i,j,k}$ are located at the cell origin at integer times.

In general, computing the discrete charge density associated with a macro-particle involves evaluating the macro-particle's assignment-function\footnote{Ref.\ \cite{hockney1988} uses symbol $W$.}
$S$.
The discrete charge density $\rho$ of a single macro-particle on the grid node $(i,j,k)$ is given by
\[
    \rho \rvert^{n}_{i,j,k} = Q_p e S(x_p - i\Delta x)S(y_p - j\Delta y)S(z_p - k\Delta z) / (\Delta x \Delta y \Delta z)
\]
where $Q_p$ is the macro-particle's charge relative to $e$, $(x_p, y_p, z_p)$ is the macro-particle's position at time step $n$, and $\Delta x$, $\Delta y$, $\Delta z$ denote the grid steps.

In detail, a trajectory is split when a macro-particle leaves its assignment cell $\tilde x_\mathrm{assignment-cell}$ \eqref{eq::assign-cell}, where $\tilde x = x/\Delta x$.
The assignment cell encloses all possible locations of a macro-particle from which charge is assigned to the same set of grid nodes in the macro-particle's neighborhood.
Accordingly, a macro-particle's assignment-cell location and boundaries depend on its position $x_\mathrm{old}$ before the movement and its assignment function $S$, respectively.
\begin{equation}\label{eq::assign-cell}
\tilde x_\mathrm{assignment-cell} \in
\begin{cases}
  [\floor(\tilde x_\mathrm{old}), \floor(\tilde x_\mathrm{old}) + 1)& \\
    \hfill \text{if assignment-function is odd order,}&\\
  [\floor(\tilde x_\mathrm{old}+0.5) - 0.5, \floor(\tilde x_\mathrm{old}+0.5) + 0.5) \qquad & \\
    \hfill \text{if assignment-function is even order,}&
\end{cases}
\end{equation}
Expressions for the assignment-cell coordinates along the $y$ and $z$ axes are similarly obtained.

The trajectory is split at the relay point, which does not need to be on the straight line path between the macro-particle's old and new position.
Several definitions for the relay point's coordinates are possible as long as the following conditions are fulfilled:
\begin{enumerate}
    \item Along the axis where the macro-particle leaves the assignment cell, the relay point's coordinate equals the assignment cell's boundary which is traversed by the macro-particle.
    \item Along the axes where the macro-particle does not leave the assignment cell, the relay point's coordinate can be anywhere within or on the boundaries of the assignment cell along this direction.
\end{enumerate}

For example, the coordinate of the relay point along the $x$-axis can be chosen as:
\begin{equation}\label{eq::relayPoint}
\tilde x_r =
\begin{cases}
  \max[\floor(\tilde x_\mathrm{old}), \floor(\tilde x_\mathrm{new})]& \\ \hfill \text{if assignment-function is odd order,}& \\
  \max[\floor(\tilde x_\mathrm{old}+0.5), \floor(\tilde x_\mathrm{new}+0.5)] - 0.5 \qquad & \\
  \hfill \text{if assignment-function is even order,}&
\end{cases}
\end{equation}
which is valid for both the macro-particle leaving the assignment cell along this axis and not leaving it.
Expressions for the relay point coordinates along the $y$ and $z$ axes are similarly obtained.

\subsection{Current Density Calculation for a Macro-Particle Staying within its Assignment Cell}
With Esirkepov's method at its core, EZ solves the electromagnetic continuity equation \eqref{eq::ECE} on the staggered Yee grid
to calculate the charge conserving single macro-particle current density $\mathbf J$ from the temporal change of the respective charge density $\rho$ \cite{yee1966}.
\begin{equation}
\label{eq::ECE}
\nabla \cdot \mathbf J + \tfrac{\partial \rho}{\partial t} = 0
\end{equation}
Within the Yee cell, the discrete current density values
\(
  \mathbf J \rvert^{n+1/2} = ( J_x \rvert^{n+1/2}_{i+1/2, j, k}, J_y \rvert^{n+1/2}_{i, j+1/2, k}, J_z \rvert^{n+1/2}_{i, j, k+1/2} )
\) are located halfway along the cell edges at times halfway between integer times, as visualized in fig.\ \ref{fig::YeeCell}.

For the derivation of the current density calculation on the grid, eq.\ \eqref{eq::ECE} is rescaled using elementary charge $e$, speed of light $c$, grid steps $\Delta x$, $\Delta y$, $\Delta z$, and the time step $\Delta t$:
\begin{itemize}
    \item The scaled charge is defined as \(
\tilde \rho = \rho \tfrac{\Delta x \Delta y \Delta z}{e}
\)
    \item The current density field is defined as \(
\mathbf {\tilde J} = \mathbf J \tfrac{\Delta x \Delta y \Delta z}{e c}
\)
\end{itemize}
Then, the electromagnetic continuity equation in finite differences reads
\begin{multline}\label{eq::ContinuityEq}
\frac{\tilde \rho \rvert^{n+1}_{i, j, k} - \tilde \rho \rvert^{n}_{i, j, k}}{c\Delta t} =
    - \frac{{\tilde J}_x \rvert^{n+1/2}_{i+1/2, j, k} - {\tilde J}_x \rvert^{n+1/2}_{i-1/2, j, k}}{\Delta x} \\
    - \frac{{\tilde J}_y \rvert^{n+1/2}_{i, j+1/2, k} - {\tilde J}_y \rvert^{n+1/2}_{i, j-1/2, k}}{\Delta y}
    - \frac{{\tilde J}_z \rvert^{n+1/2}_{i, j, k+1/2} - {\tilde J}_z \rvert^{n+1/2}_{i, j, k-1/2}}{\Delta z}\,.
\end{multline}
Since the charge densities $\tilde \rho \rvert^{n}_{i, j, k}$ and $\tilde \rho \rvert^{n+1}_{i, j, k}$ are known from the macro-particle's position before and after movement, respectively, $\tilde{\mathbf J}\rvert^{n+1/2}$ can be calculated from \eqref{eq::ContinuityEq}.
Only current densities associated with the movement of a single macro-particle need to be considered, because the total current density on a grid node is obtained by summing current densities on that grid node of all macro-particles.

A charge-conserving current density of a single macro-particle is obtained from the continuity equation\ \eqref{eq::ContinuityEq} by inserting this macro-particle's charge density before and after a movement.
Let the macro-particle's position at time $t_n=n\Delta t$ before a movement be $(x_\mathrm{old}, y_\mathrm{old}, z_\mathrm{old})$ and at time $t_{n+1}=(n+1)\Delta t$ after a movement be $(x_\mathrm{new}, y_\mathrm{new}, z_\mathrm{new})$.
Denoting the macro-particle's charge as $Q$, and abbreviating $S(x-i\Delta x) \eqdef S_i(x)$ ($y$ and $z$ direction similarly), the scaled single macro-particle current density is related to the current deposition vector $\mathbf W \rvert^{n+1/2}_{i,j,k}$:
\gl{\label{eq::WDefinition}
\begin{split}
    {\tilde J}_x \rvert^{n+1/2}_{i+1/2, j, k} - {\tilde J}_x \rvert^{n+1/2}_{i-1/2, j, k}& = -Q \frac{\Delta x}{c \Delta t} W_x \rvert^{n+1/2}_{i,j,k} \\
    {\tilde J}_y \rvert^{n+1/2}_{i, j+1/2, k} - {\tilde J}_y \rvert^{n+1/2}_{i, j-1/2, k}& = -Q \frac{{\Delta y}}{c \Delta t} W_y \rvert^{n+1/2}_{i,j,k} \\
    {\tilde J}_z \rvert^{n+1/2}_{i, j, k+1/2} - {\tilde J}_z \rvert^{n+1/2}_{i, j, k-1/2}& = -Q \frac{{\Delta z}}{c \Delta t} W_z \rvert^{n+1/2}_{i,j,k}\,,
\end{split}
}
with the current deposition vector $\mathbf W \rvert^{n+1/2}_{i,j,k}$ being defined from the continuity eq.\ \eqref{eq::ContinuityEq} for the movement of a single macro-particle:
\begin{multline*}
S_i(x_\mathrm{new})S_j(y_\mathrm{new})S_k(z_\mathrm{new})
  - S_i(x_\mathrm{old})S_j(y_\mathrm{old})S_k(z_\mathrm{old}) \\
= W_x \rvert^{n+1/2}_{i,j,k} + W_y \rvert^{n+1/2}_{i,j,k} + W_z \rvert^{n+1/2}_{i,j,k}\,.
\end{multline*}

Assuming the macro-particle travels along a straight line by a distance not more than one cell and following the notation established by Esirkepov \cite{esirkepov2001}, the components of $\mathbf W$ are related to the assignment-function of the macro-particle as follows.
Defining the helper function
\begin{multline*}
W(s_1, s_2, s_3, s_4, s_5, s_6) = \\
\frac{1}{3} \left(
  s_4 s_5 s_6
  - s_1 s_5 s_6
  + s_4 s_2 s_3
  - s_1 s_2 s_3
\right)
\\
+ \frac{1}{6} \left(
  s_4 s_2 s_6
  - s_1 s_2 s_6
  + s_4 s_5 s_3
  - s_1 s_5 s_3
\right)\,,
\end{multline*}
the components of $\mathbf W$ can be calculated by
\gl{\label{eq::WCalculation}
\begin{split}
& W_x \rvert^{n+1/2}_{i,j,k} = \\
  & \quad W(S_i(x_\mathrm{old}), S_j(y_\mathrm{old}), S_k(z_\mathrm{old}), S_i(x_\mathrm{new}), S_j(y_\mathrm{new}), S_k(z_\mathrm{new})) \\
& W_y \rvert^{n+1/2}_{i,j,k} = \\
  & \quad W(S_j(y_\mathrm{old}), S_i(x_\mathrm{old}), S_k(z_\mathrm{old}), S_j(y_\mathrm{new}), S_i(x_\mathrm{new}), S_k(z_\mathrm{new})) \\
& W_z \rvert^{n+1/2}_{i,j,k} = \\
  & \quad W(S_k(z_\mathrm{old}), S_j(y_\mathrm{old}), S_i(x_\mathrm{old}), S_k(z_\mathrm{new}), S_j(y_\mathrm{new}), S_i(x_\mathrm{new}))\,.
\end{split}
}

Current deposition by Esirkepov's method generally takes the macro-particle's location before and after movement as `old' and `new' coordinates.
EZ makes the same choice as long as the macro-particle does not leave its assignment cell during movement.
This already yields a current density for Esirkepov's method and EZ which is different from ZigZag's, as is shown in the supplemental material for macro-particle assignment-function CIC.

\subsection{Current Density Calculation for a Macro-Particle Leaving its Assignment Cell}
If the macro-particle leaves its assignment cell during movement, EZ makes a different choice for the `old' and `new' coordinates compared to Esirkepov's method.
EZ takes inspiration from the ZigZag approach of splitting the trajectory in two parts.
The total current density associated with the macro-particle movement is then obtained by summing the contributions of two virtual macro-particles, each virtual macro-particle traveling along its respective part of the split trajectory during the whole time step.
That is, the `old' and `new' position of the first virtual macro-particle refer to the original macro-particle's initial position and the coordinates of the relay point, while they refer to the relay point and the original macro-particle's final position for the second virtual macro-particle.
Since each of the virtual macro-particle movements is restricted to the respective virtual macro-particle's assignment cell, the associated single virtual macro-particle current density can be calculated by Esirkepov's method as outlined above.

The procedure of splitting the trajectory at the relay point and adding up the current densities of two virtual macro-particles preserves the finite-difference continuity equation.
Charge density value $\tilde \rho^\prime \rvert^{n+1}$ introduced at time $t_{n+1}$ by the first virtual macro-particle located at the relay point and charge density value $\tilde \rho^\prime \rvert^{n}$ introduced at time $t_{n}$ by the second virtual macro-particle located at the relay point are equal and together represent a static charge density distribution which does not generate a current.

Namely, the current density values produced by the virtual macro-particles separately adhere to their respective continuity equations with respective left-hand sides:
\begin{equation*}
\tfrac{\tilde \rho^\prime \rvert^{n+1}_{i,j,k} - \tilde \rho \rvert^{n}_{i,j,k}}{c\Delta t} \text{ and } \tfrac{\tilde \rho \rvert^{n+1}_{i,j,k} - \tilde \rho^\prime \rvert^{n}_{i,j,k}}{c\Delta t}
\end{equation*}

Summing up these equations and taking into account $\tilde \rho^\prime \rvert^{n+1} = \tilde \rho^\prime \rvert^{n}$ yields eq.\ \eqref{eq::ContinuityEq} for the combined current density.
Thus, EZ current deposition is charge-conserving.

However, due to discretization, the implementation-specific choice of the relay point can result in different values for the components of the current deposition vector $\mathbf W$ and therefore different grid values of $\mathbf J$ for the farthest non-vanishing grid nodes.
We emphasize that the finite-difference continuity eq.\ \eqref{eq::ContinuityEq} is preserved with any of the relay point choices (satisfying the conditions).
This fact has already been pointed out in the original article \cite{esirkepov2001} with its uniqueness lemma being tied to the straightforward trajectory assumption.

Since numerical dispersion, as well as other particle-in-cell simulation properties, depend on the current deposition algorithm \cite{godfrey2013, meyers2015, xiao2019}, it is possible that certain choices of the relay point's coordinates yield favorable numerical conditions.

In case of 2D3V simulations, where all fields and assignment-functions depend only on $x$ and $y$, the current density components $J_x$ and $J_y$ are calculated using the same trajectory splitting approach for movements along $x, y$.
The corresponding components $W_x$ and $W_y$ of the current deposition vector are calculated from equation \ref{eq::WCalculation} with $S_k(z) = 1$ for all $z$.
After simplification, it yields:
\begin{equation*}
\begin{split}
W_x \rvert^{n+1/2}_{i,j,k}& = W(S_i(x_\mathrm{old}), S_j(y_\mathrm{old}), 1, S_i(x_\mathrm{new}), S_j(y_\mathrm{new}), 1) \\
  & = \frac{1}{2} \left( S_i(x_\mathrm{new})S_j(y_\mathrm{new}) - S_i(x_\mathrm{old})S_j(y_\mathrm{new}) \right. \\
  & \quad + \left. S_i(x_\mathrm{new})S_j(y_\mathrm{old}) - S_i(x_\mathrm{old})S_j(y_\mathrm{old}) \right) \\
W_y \rvert^{n+1/2}_{i,j,k}& = W(S_j(y_\mathrm{old}), S_i(x_\mathrm{old}), 1, S_j(y_\mathrm{new}), S_i(x_\mathrm{new}), 1)\\
  & = \frac{1}{2} \left( S_i(x_\mathrm{new})S_j(y_\mathrm{new}) - S_i(x_\mathrm{new})S_j(y_\mathrm{old}) \right. \\
  & \quad + \left. S_i(x_\mathrm{old})S_j(y_\mathrm{new}) - S_i(x_\mathrm{old})S_j(y_\mathrm{old}) \right)\,.
\end{split}
\end{equation*}

However, for movement along $z$ there is no grid or coordinate change and therefore no trajectory splitting is possible.
In this case we use the same way of calculating $J_z$ as eq.\ (37) in \cite{esirkepov2001}, which in our notation becomes
\begin{multline}
\label{eq::Jz2d}
{\tilde J}_z \rvert^{n+1/2}_{i, j, k+1/2} = \\
  Q \frac{v_z}{c} \left( \frac{1}{3} S_i(x_\mathrm{new})S_j(y_\mathrm{new}) + \frac{1}{6} S_i(x_\mathrm{old})S_j(y_\mathrm{new})\right. \\
  + \left. \frac{1}{6} S_i(x_\mathrm{new})S_j(y_\mathrm{old}) + \frac{1}{3} S_i(x_\mathrm{old})S_j(y_\mathrm{old}) \right)
\end{multline}
where $v_z$ is velocity along $z$ at time $t_{n+1/2}$.

\section{Efficiency Increase of EZ Compared to Esirkepov's Method}
EZ can significantly shorten the calculation time of the current density field compared to Esirkepov's method.
This originates from a reduction of the necessary computations of the current deposition vector $\mathbf W$ as explained in the following.

\begin{figure}[tb]
  \centering
  \includegraphics[width=0.45\textwidth]{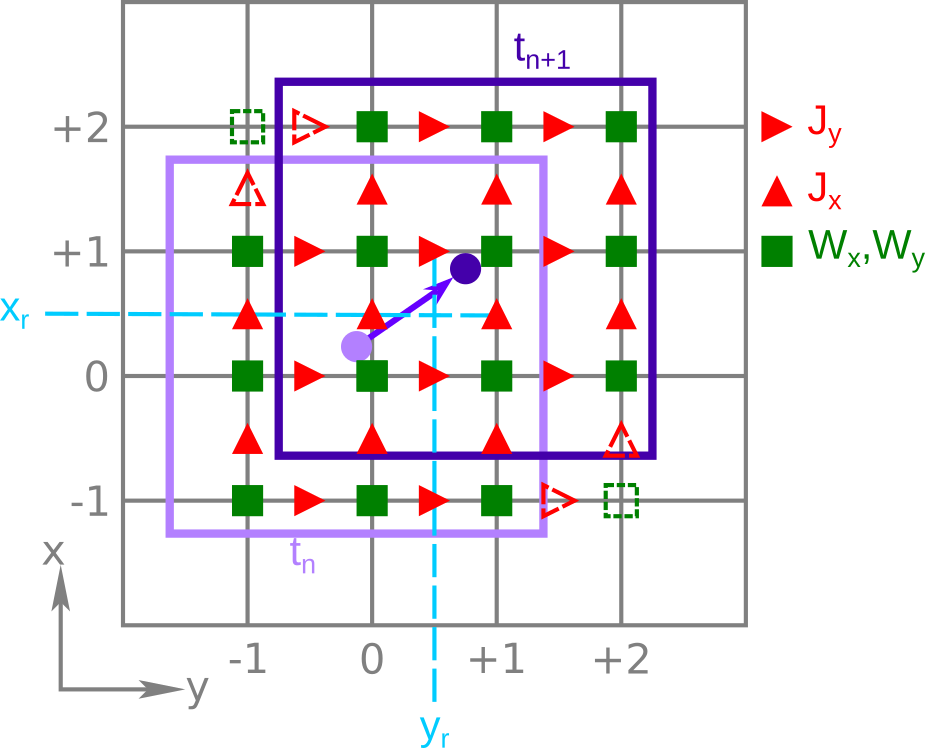}
  \caption{
    Visualization of discrete current deposition on two-dimensional grid for Esirkepov's method with second order assignment function TSC.
  }
  \label{fig::EsirkepovCurrentDeposition}
\end{figure}

For a macro-particle leaving its assignment cell during step from time $t_n$ to $t_{n+1}$, fig.\ \ref{fig::EsirkepovCurrentDeposition} visualizes the non-vanishing entries of the current deposition vector components $W_x$, $W_y$, as well as current density $J_x$, $J_y$ on a 2D grid for Esirkepov's method.
The relay point $(x_r, y_r)$ marks the position where charge assignment extends over a different set of grid nodes when crossed by the macro-particle.
The extent of the macro-particle's assignment-function at times $t_n$ and $t_{n+1}$ are highlighted for the second-order assignment-function TSC \cite{hockney1988}.
The number of nodes with non-vanishing values of $\mathbf W$ and $\mathbf J$ is within the bounded area.
Albeit the values of the dashed entries are zero in the special case of this drawing, they are non-zero for a movement in another direction and therefore need to be calculated in a generally applicable implementation of Esirkepov's method.
However, the last row (column) of $W_x$ ($W_y$) at $x=+2$ ($y=+2$) does not need to be calculated, since the next row (column) of this quantity along $+x$ ($+y$) is zero and therefore the row (column) of current density $J_x$ ($J_y$) at $x=+2+1/2$ ($y=+2+1/2$) vanishes according to eq.\ \eqref{eq::WDefinition}.
For the general 2D case, there are in total $4\times3=12$ evaluations of each $W_x$ and $W_y$ required to calculate the respective current density component with Esirkepov's original method.
See eq.\ \eqref{eq::Jz2d} for the additional calculation of $J_z$ that needs to be performed in 2D3V simulations.

In general, for an assignment-function of order $l$, where charge is assigned to $(l+1)^d$ points in a $d$-dimensional simulation, this requires a maximum of
\(
(l+2)^{d-1} \times (l+1)
\)
evaluations of a component of the current deposition vector $\mathbf W$ per macro-particle with Esirkepov's original method.
This assumes the most general case where the nodes, to which a macro-particle's charge is assigned, change along every axis due to its movement.

\begin{figure*}[t]
 \centering
 \subfloat{\includegraphics[width=0.40\textwidth]{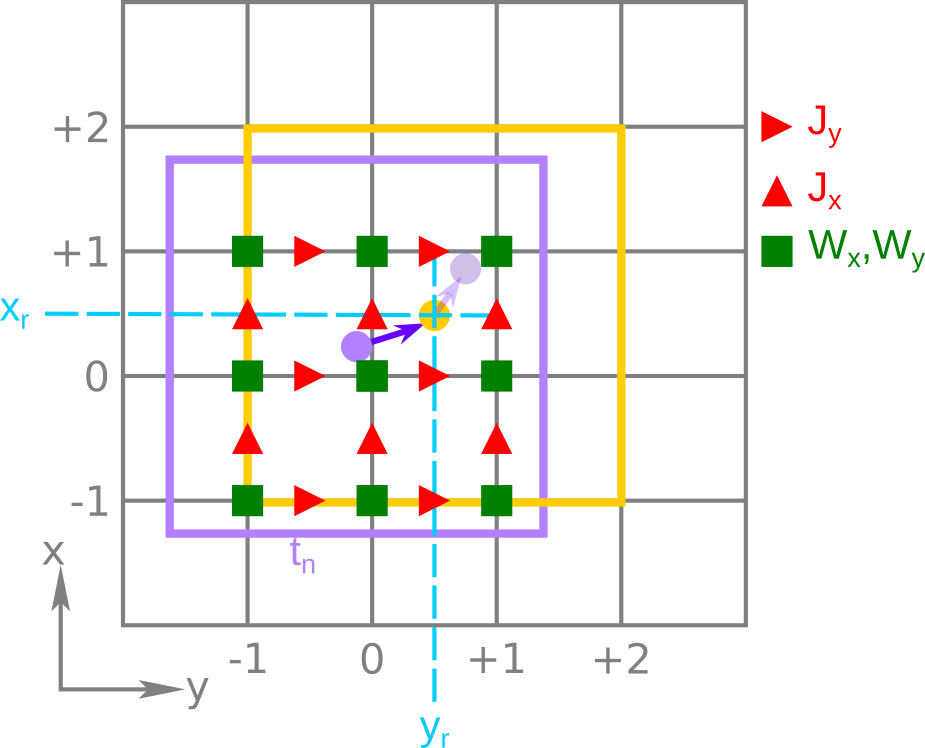}}
 \hfil
 \subfloat{\includegraphics[width=0.40\textwidth]{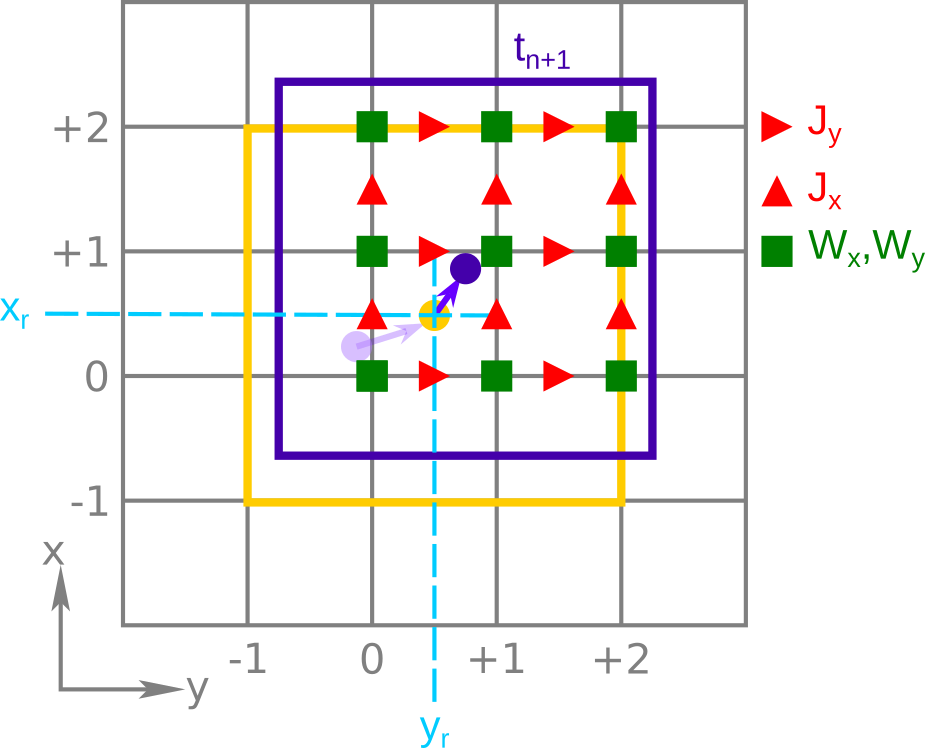}}
 \caption{
  EZ-specific visualization of two-dimensional grid and second order assignment function TSC.
  The left hand side pictures the first part while the right hand side pictures the second part of the split trajectory.}
 \label{fig::EzCurrentDeposition}
\end{figure*}

For EZ, fig.\ \ref{fig::EzCurrentDeposition} visualizes the non-vanishing entries of $W_x$, $W_y, J_x$, $J_y$ similarly to fig.\ \ref{fig::EsirkepovCurrentDeposition}.
Two independent but equal virtual macro-particles each travel along one part of the split trajectory.
By the same argument as used for the classic method by Esirkepov, $3\times2=6$ evaluations of each $W_x$ and $W_y$ are required per part of the trajectory.
Although this is the same number as for Esirkepov's method (in this 2D example), EZ offers improved performance compared to direct application of Esirkepov's method.

Esirkepov's traditional method must calculate the current density on all nodes to which charge of the macro-particle may be assigned before and after movement.
This includes evaluations of the assignment-function out of its support making the assignment function implementation more costly compared to EZ.
Due to path splitting in EZ, assignment function evaluations are always restricted to its support.
While this reduces the complexity of assignment function implementations on the one hand, it also stimulates parallel calculation of the current density from many macro-particles by equalizing between all macro-particles the number of nodes at which the single contributions are calculated.
That is, there is a fixed set of operations performed equally for all macro-particles in the current density calculation without branching from the instruction sequence.

Moreover, in a $d$-dimensional simulation EZ reduces evaluations of the current deposition vector components to
$
2 \times (l+1)^{d-1} \times l
$
points, as it spares calculations where the current density evaluates to zero.
In 3D, these are fewer than the required amount of points for Esirkepov's method for assignment-functions of order $l=3$ (PQS) and lower.

In summary, employing the trajectory splitting in principle allows EZ to surpass the performance of Esirkepov's method for two reasons.
\begin{enumerate}
  \item EZ evaluates the current deposition vector $\mathbf W$ at fewer positions for lower order assignment-functions.
  \item Reduced assignment-function definitions can be utilized where evaluations at positions outside the support do not need to be handled since all assignment-function evaluations are on-support with EZ.
\end{enumerate}
As explained above, the latter equalizes the instruction sequence for all macro-particles which keeps thread branch divergence low, thereby promoting parallel computation via SIMT (single instruction, multiple threads) while avoiding serialization and idling threads.
This is especially important when running on GPUs.
Moreover, with the support size being a fixed value known at compile time, compilers may be able to optimize loops in the calculation of $\mathbf W$.
On CPUS, these loops can be auto-vectorized promoting parallel computation via SIMD (single instruction, multiple data).
Specifically for PQS, where the number of nodes with non-zero entries of the current deposition vector components is four in two directions, iterations of the $\mathbf W$ calculation loop and writes of the result to memory can be directly expressed with SSE4 (Streaming SIMD Extensions 4) intrinsics for single precision or AVX2 (Advanced Vector Extensions 2) intrinsics for double precision, explicitly enforcing usage of SIMD operations instead of relying on auto-vectorization.

\section{EZ Implementation in PIConGPU}
\label{sec::EzImplementationDetails}

EZ's implementation in PIConGPU had the design goals:
\begin{enumerate}
    \item Keep register footprint low to increase (GPU) occupancy \cite{occupancy2023}.
    \item Keep memory accesses, especially writes, low to avoid latency.
\end{enumerate}

To address goal (1), the calculation of every component of the current deposition vector $\mathbf W$ is performed independently.
This results in multiple calculations of the same assignment-function value instead of storing it in register memory, accomplishing the first goal.

Goal (2) provides the reason for the choice of the relay point.
The initial coordinates of the relay point are the coordinates of the macro-particle's location after movement.
Only if a macro-particle leaves its assignment cell along one axis, is the coordinate of the relay point on this axis changed.
The relay point coordinates along the other axes remain unchanged as long as the macro-particle does not leave the assignment cell along these axes too.
This allows cutting off calculations of whole components of the current density unless necessary.

For example, if the macro-particle is still within its assignment cell after particle push, the first virtual particle will move along the whole trajectory while the second virtual particle will not move at all.
Therefore, the second virtual particle's current density calculation is completely spared in PIConGPU.
While this possibly leads to branching of threads once per macro-particle, the performance impact is not as significant as branching during node level calculation of the single macro-particle current density.

Next, consider a macro-particle leaving its assignment cell only along one axis, say $x$.
The relay point coordinates are at the assignment cell boundary along $x$ and at the macro-particle's final position along $y$.
Then the first virtual particle changes its position along $x$ and $y$ while it moves from the macro-particle's initial position to the relay point.
The second virtual particle moving from the relay point to the macro-particle's final position then only travels parallel to the $x$-axis allowing to spare the computation of the $y$-component of its associated current density as it is zero.
Moreover, a computation of the $z$ component of the current density is not performed for either virtual particle.

\begin{table*}[t]
\caption{Comparison of the number of nodes with non-zero entries of the current deposition vector for single macro-particle current deposition by Esirkepov's method and EZ for a second-order (TSC) macro-particle assignment-function.}
\centering
\tiny
\begin{tabular}{| r | c c c | c c c | c | c |}
\hline
\multicolumn{1}{|c|}{macro-particle}
    & \multicolumn{6}{|c|}{number of nodes with non-zero entries of the current deposition vector component}
    & \multicolumn{2}{|c|}{total writes to} \\
\multicolumn{1}{|c|}{leaves along}
    & \multicolumn{3}{|c|}{Esirkepov}
    & \multicolumn{3}{|c|}{EZ}
    & \multicolumn{2}{|c|}{current density array} \\
\multicolumn{1}{|c|}{direction} & $W_z$ & $W_y$ & $W_x$ & $W_z$ & $W_y$ & $W_x$ & Esirkepov & EZ \\
\hline
none & $2 \times 3 \times 3$ & $3 \times 2 \times 3$ & $3 \times 3 \times 2$
    & $2 \times 3 \times 3$ & $3 \times 2 \times 3$ & $3 \times 3 \times 2$
    & 54 & 54\\
$x$ & $2 \times 3 \times 4$ & $3 \times 2 \times 4$ & $3 \times 3 \times 3$
    & $2 \times 3 \times 3$ & $3 \times 2 \times 3$ & $2\times(3 \times 3 \times 2)$
    & 75 & 72\\
$y$, $x$ & $2 \times 4 \times 4$ & $3 \times 3 \times 4$ & $3 \times 4 \times 3$
     & $2 \times 3 \times 3$ & $2\times(3 \times 2 \times 3)$ & $2\times(3 \times 3 \times 2)$
     & 104 & 90\\
$z$, $y$, $x$ & $3 \times 4 \times 4$ & $4 \times 3 \times 4$ & $4 \times 4 \times 3$
    & $2\times(2 \times 3 \times 3)$ & $2\times(3 \times 2 \times 3)$ & $2\times(3 \times 3 \times 2)$
    & 132 & 108\\
\hline
\end{tabular}
\label{tab::EsirkepovEzOperations}
\end{table*}

Table \ \ref{tab::EsirkepovEzOperations} compares the number of nodes with non-zero entries of the components of the current deposition vector and the resulting total number of writes to the current density array between Esirkepov's method and EZ.
A second-order (TSC) macro-particle assignment-function \cite{hockney1988} is considered as an example.
The number of nodes with non-zero entries, and of current density writes accordingly, depend on the number of directions along which the macro-particle leaves its assignment cell.
In the given order, the cases correspond to not leaving the cell, or leaving along one, two, or three directions, respectively.
The number of total writes given in the table is for a macro-particle moving along all three spatial directions, such that all current deposition vector components are non-zero.
The total number of writes equals the number of nodes with non-zero current deposition vector entries summed over all components.
A lower number of calculations and writes presumably results in better performance.
The number of nodes with non-zero entries is given per dimension, where the order corresponds to $(z, y, x)$.

Note, in this comparison an optimization of Esirkepov's method has been exploited as follows.
If a macro-particle does not leave its assignment cell along all directions, there are nodes at which computations of the current-distribution vector can be spared, as values at these nodes are zero.
Sparing these computations reduces computational time as well as memory writes.
These nodes are oriented along the directions where the macro-particle stays within its assignment cell.
That is, if a macro-particle moves in a specific direction, but stays within its assignment cell along this direction, the number of nodes at which the current deposition vector components need to be evaluated are reduced by one along this direction.
PIConGPU applies these optimizations in its implementation of Esirkepov's method, which is relevant for the performance comparisons in sec.\ \ref{sec::PerformanceComparison}.

As can be seen in tab.\ \ \ref{tab::EsirkepovEzOperations}, EZ outperforms Esirkepov's method in all cases but the first one where they are equal.
In this first case, they are equal due to the above described optimization applied to Esirkepov's method.
Since optimized assignment-function implementations can be used for EZ, it can still be faster than Esirkepov's method for this case in practice.
Overall, EZ shows potential for improving Esirkepov's performance.

Lower order assignment-functions, e.\,g.\ zeroth-order (NGP) or first-order (CIC), increase the difference between the Esirkepov's method and EZ with EZ requiring fewer calculations and writes.

In cases where assignment functions of fourth-order and higher are required to minimize aliasing effects, EZ's favorable properties of reduced assignment-function complexity and uniform instruction sequence can offer a benefit compared to Esirkepov's method.

\section{Physics Acceptance Tests and Performance Results}

\subsection{Level of Charge Conservation}
\subsubsection{Single Macro-Particle Simulation}
In order to evaluate EZ's level of charge conservation and to compare this to the optimized Esirkepov's method and ZigZag, a simple single moving macro-particle test case is set up in PIConGPU.

The test consists of one macro-particle of first-order assignment-function CIC \cite{hockney1988} with charge $-e$ in a $24\times24\times24$ cells simulation volume.
At the beginning it is located at $\mathbf r = (8.9\Delta x, 8.8\Delta y, 8.7\Delta z)$ and moving with constant normalized speed $v/c = \beta = 0.999$.
The test comprises three scenarios, each characterized by the number of directions along which the macro-particle leaves its assignment cell.
It does so either along the $x$ axis only, within the $x$ and $y$ plane only, or diagonally across the volume.
The respective velocity vectors are $\mathbf v = v \cdot (1, 0, 0)$, $\mathbf v = v \cdot (1, 1, 0) / \sqrt{2}$, or $\mathbf v = v \cdot (1, 1, 1) / \sqrt{3}$.

Within the PIC iteration step, the electric field $\mathbf E$ and charge density field $\rho$ are calculated by PIConGPU.
These are written to disk after one time step $\Delta t = 0.5 \frac{\Delta x}{c}$ and then used to calculate the normalized remainder $\tilde R_\mathrm{SMP}$ of Gauss's law.
That is, the field $R\rvert^n = \mathbf{\nabla} \cdot \mathbf E\rvert^n - \rho^\prime\rvert^n$ normalized to elementary charge per cell volume $\rho_\mathrm{SMP}=e/(\Delta x \Delta y \Delta z)$ such that $\tilde R_\mathrm{SMP} = R / \rho_\mathrm{SMP}$.
The charge density $\rho^\prime\rvert^n = \rho\rvert^n - \rho\rvert^0$ used in the calculation of $R\rvert^n$ is a composition of the single macro-particle charge density $\rho\rvert^n$ at time step $n$ and the sign-inverted single macro-particle charge density at simulation start $-\rho\rvert^0$, in order to account for the virtual mirror charge on the grid created by the finite-difference time-domain electromagnetic field solver employed by PIConGPU, as it assumes zero field at simulation startup.
$\tilde R_\mathrm{SMP}$ quantifies the level of charge conservation in the simulations.
Ideally, $\tilde R_\mathrm{SMP}$ is zero in case of perfect charge conservation and absent floating point round-off errors.

For this single macro-particle test case, the level of charge conservation $\lambda_\mathrm{SMP}$ is quantified by the largest absolute value of $\tilde R_\mathrm{SMP}$ on the grid.
That is, $\lambda_\mathrm{SMP} = \|\tilde R_\mathrm{SMP}\|_\infty \defeq \max\{|\tilde R_\mathrm{SMP} \rvert_{ijk}|\}$.
Results are displayed in Table \ \ref{tab::ChargeConservationSimulation}.
The source codes of simulations and analysis are available online \cite{widera2022}.

\begin{table}[t]
\caption{Maximum absolute value $\lambda_\mathrm{SMP}$ of the normalized remainder $\tilde R_\mathrm{SMP}$ of Gauss's law for simulations of a single moving macro-particle with different current deposition methods.
}
\tiny
\centering
\begin{tabular}{| r | c | c | c |}
\hline
\multicolumn{1}{|c|}{macro-particle}
    & \multicolumn{3}{|c|}{level of charge conservation $\lambda_\mathrm{SMP}$} \\
\multicolumn{1}{|c|}{leaves along}
    & optimized Esirkepov's method & ZigZag & EZ \\
\hline
$x$ & 3.6e-08 & 3.6e-08 & 3.6e-08 \\
$y$, $x$ & 5.9e-08 & 4.1e-08 & 4.1e-08 \\
$z$, $y$, $x$ & 5.8e-08 & 1.3e-3 & 5.8e-08 \\
\hline
\end{tabular}
\label{tab::ChargeConservationSimulation}
\end{table}

Esirkepov's method and EZ reach comparable charge conservation in all three scenarios, where the absolute level of charge conservation is on the order of rounding error of floating-point arithmetic operations.
ZigZag is on the same level as Esirkepov's method, as long as the macro-particle does not move diagonally through the volume.
In the case of the macro-particle moving diagonally through the volume, ZigZag's charge conservation is orders of magnitude worse compared to Esirkepov's method and EZ.
We suspect, the 3D eqs.\ (20) of ZigZag in \cite{umeda2003} are actually only valid in 2D as they miss contributions to the current density $J$ as in eqs.\ (35)-(37) in \cite{villasenor1992}, which are proportional to the product of a macro-particle's displacements in all three dimensions.
Due to this shortcoming of ZigZag, we will not take it into account in the following comparisons.

\subsubsection{Warm Plasma Simulation}
In order to evaluate EZ's level of charge conservation in a real world simulation and to compare it to the optimized Esirkepov's method, we set up a warm plasma simulation.
The simulation consists of an initially field free space in which there are approximately 177 million electron-like macro-particles moving at a relativistic velocity.
Due to macro-particle speeds close to the speed of light, we anticipate about half of the macro-particles to leave their assignment cell per step which is sufficient to compare charge conservation of EZ and Esirkepov's method.

The simulation domain extends over 192 cells in each direction and employs periodic boundary conditions on each side.
There are 25 macro-particles per cell which sample a homogeneous distribution of relativistic electrons at a density of $\rho_\mathrm{WP} = - 10^{20}\,e /\mathrm{m}^3$.
Their initial momenta are normally distributed along each spatial dimension with a variance of $17.5\,m^2c^2$, resulting in a distribution of initial total momentum according to a Maxwell-Boltzmann distribution with a temperature of 8942\,keV and a most probable kinetic energy of $5\,mc^2$.
The grid resolution of $\Delta x = 57.8918$\,{\textmu}m is equal along all directions and chosen such that the simulation volume encompasses 5 Debye lengths along one axis at the given density and initial temperature.
Time resolution is $\Delta t = 0.5 \Delta x / c$, providing variability with respect to the choice of Maxwell solver.
The simulation setup and analysis code are available online \cite{widera2022}.

Whereas a local estimate for the relative error of charge conservation at one time step can be defined by the normalized remainder $\tilde R_\mathrm{WP} = R/\rho_\mathrm{WP}$ of Gauss' law at each grid node,
a global estimate for the relative error of charge conservation at one time step can be obtained by analyzing the scattering of all $\tilde R_\mathrm{WP}$ values on the grid around the expectation value zero.
We quantify the global charge conservation level $\lambda_\mathrm{WP}$ by the estimated standard deviation of the $\tilde R_\mathrm{WP}$ sample
\gl{\label{eq::chargeConservationWP}
  \lambda_\mathrm{WP} = \sqrt{ \frac{1}{N_\mathrm{grid}} \sum_{l=1}^{N_\mathrm{grid}} {\tilde{R}_{\mathrm{WP},l}^2}}\,,
}
where $N_\mathrm{grid}$ is the total number of $\tilde R_\mathrm{WP}$ values at one time step, i.\,e.\ $(192-1)^3$.

\begin{figure}[t]
 \centering
 \includegraphics[width=0.45\textwidth]{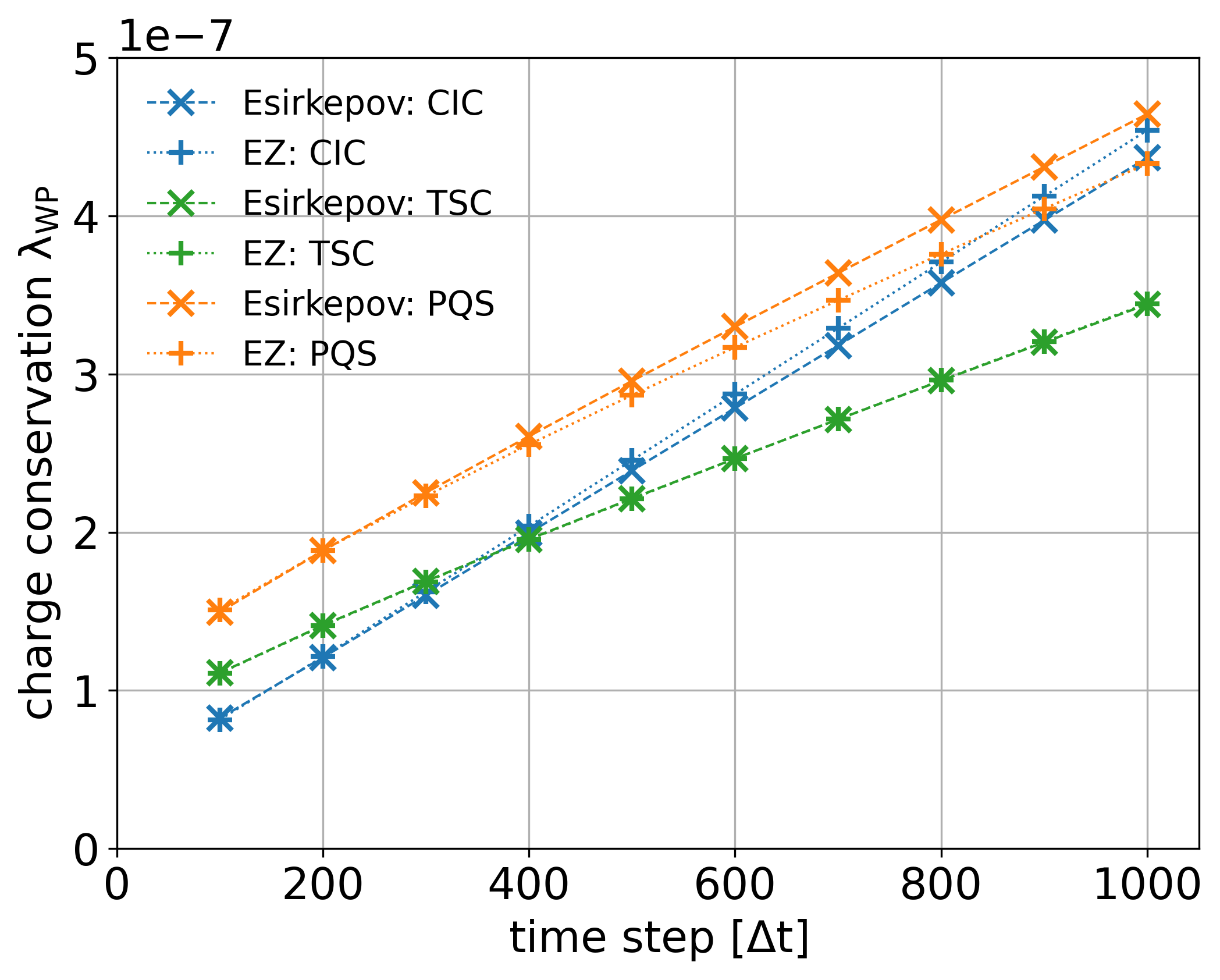}
 \caption{
    Evolution of charge conservation over time in the warm plasma simulation for EZ and Esirkepov's method and different macro-particle assignment-functions of increasing order.
    Charge conservation $\lambda_\mathrm{WP}$ is quantified by the standard deviation of the normalized Gauss' law remainder $\tilde R_\mathrm{WP}$ distribution on the grid, eq.\ \eqref{eq::chargeConservationWP}.
 }
 \label{fig::WarmPlasmaChargeConservation}
\end{figure}

Evolution of global charge conservation $\lambda_\mathrm{WP}$ for EZ and Esirkepov's method during the course of simulation is depicted in Fig.\ \ref{fig::WarmPlasmaChargeConservation} for the three macro-particle assignment-functions first-order CIC, second-order TSC, and third-order PQS \cite{hockney1988}.
Global charge conservation is on the same order of magnitude for both EZ and Esirkepov's method for all three macro-particle assignment-functions.
At this order of magnitude a scattering of global charge conservation values is expected due to numerical floating point round off errors in parallel atomic summation of single macro-particle currents in the current deposition algorithm.

The uncertainty of $\lambda_\mathrm{WP}$ is estimated to be $\lambda_\mathrm{WP}/\sqrt{2n}$ \cite{zyla2020}.
For all data points in Fig.\ \ref{fig::WarmPlasmaChargeConservation}, this is below or equal to $1.2\times10^{-10}$ which happens to be smaller than the markers of the data points and therefore is not visible.

From the results of the single macro-particle and the warm plasma simulation we conclude successful validation of EZ.

\subsection{Performance Comparison}
\label{sec::PerformanceComparison}

In order to examine the expected scaling of run time with macro-particle assignment-function order for the optimized Esirkepov's method and EZ, we run the above warm plasma simulation setup for several macro-particle assignment-functions and on two different GPUs, as well as on a CPU.
Recall that optimizations applied to Esirkepov's method are laid out in Sec.\ \ref{sec::EzImplementationDetails}.

\subsubsection{Setup}
Simulations ran for assignment-functions from first-order CIC to third-order PQS \cite{hockney1988} on one NVIDIA Tesla V100 16GiB GPU at Summit \cite{summitWWW}, one AMD Instinct MI100 GPU at Spock \cite{spockWWW}, and one 64-core AMD EPYC 7662 CPU at Spock \cite{spockWWW}.

NVIDIA V100 simulations on Summit's IBM Power9 nodes were compiled with \texttt{nvcc} for architecture \texttt{sm\_70} to target \texttt{NVIDIA V100} with \texttt{CUDA 11.0.3}, \texttt{spectrum-mpi 10.4.0.3-20210112}, and \texttt{boost 1.74.0} in the software environment.

AMD MI100 simulations on Spock's AMD EPYC 7662 64-core nodes were compiled with \texttt{hipcc} for architecture \texttt{gfx908} to target \texttt{AMD MI100} with \texttt{ROCm 4.5.0}, \texttt{cray-mpich 8.1.10}, and \texttt{boost 1.73.0} in the software environment.

AMD EPYC 7662 64-core simulations were compiled with \texttt{CCE Clang 12.0.3} with \texttt{-march=native} on the compute node with \texttt{cray-mpich 8.1.10}, as well as \texttt{boost 1.73.0} in the software environment and executed with 128 threads.

For each assignment-function and processing unit, ten simulations executing one hundred time steps were performed.
Only the run time of the simulation steps was measured; simulation initialization and finalization are not included in the measured time.

Since current deposition is just one part of a particle-in-cell iteration step, whose fraction of time taken in the total simulation step increases with higher macro-particle assignment-function order, we additionally profiled the kernel on the different GPU architectures.
Profiling allowed us to quantify the impact of EZ on the current deposition kernel alone.
Profiles were taken for the first five time steps of the warm plasma simulation.
After an initial warm-up phase of two steps, profiles did not change significantly in the following 3 steps.
Profiling results presented in the following are taken from time step five.

All simulations for one compute architecture ran successively on the same node to avoid confounding hardware performance differences between nodes.
GPU simulations were performed on a single GPU in single precision.

\subsubsection{Results}
\begin{table*}[t]
\caption{
Performance metrics of EZ, as well as optimized Esirkepov's current deposition (CD) method for the warm plasma simulation with respect to macro-particle assignment-functions.
The ratio of time taken by optimized Esirkepov's method over EZ defines speedup.
}
\centering
\begin{tabular}{| c | c | c | c | c | c | c |}
\hline
Assignment
    & \multicolumn{3}{|c|}{CD kernel run time [ms]}
    & \multicolumn{3}{|c|}{Avg. time per step [ms]}
    \\
function
    & Esirkepov & EZ & speedup
    & Esirkepov & EZ & speedup
    \\
\hline
\multicolumn{7}{|c|}{NVIDIA V100 16GiB}\\
\hline
CIC  & 31.30 & 21.68 & 1.444
  & 88.15 $\pm$ 0.07 & 80.04 $\pm$ 0.06 & 1.100 \\
TSC  & 109.72 & 94.69 & 1.159
  & 164.24 $\pm$ 0.06 & 151.65 $\pm$ 0.04 & 1.083 \\
PQS & 297.33 & 252.95 & 1.175
  & 356.16 $\pm$ 0.03 & 319.91 $\pm$ 0.06 & 1.113 \\
\hline
\multicolumn{7}{|c|}{AMD MI100}\\
\hline
CIC & 20.81 & 16.55 & 1.257
  & 144.66 $\pm$ 0.52 & 139.43 $\pm$ 0.79 & 1.034 \\
TSC  & 79.56 & 70.49 & 1.129
  & 226.93 $\pm$ 0.79 & 217.02 $\pm$ 0.73 & 1.046 \\
PQS  & 179.87 & 182.95 & 0.983
  & 363.83 $\pm$ 0.80 & 366.70 $\pm$ 0.57 & 0.992 \\
\hline
\multicolumn{7}{|c|}{AMD EPYC 7662 64-core}\\
\hline
CIC & \multicolumn{3}{|c|}{n/a} & 892.1 $\pm$ 0.4 & 864.9 $\pm$ 0.4 & 1.031 \\
TSC  & \multicolumn{3}{|c|}{n/a} & 1904.3 $\pm$ 0.5 & 1674.2 $\pm$ 2.1 & 1.137 \\
PQS & \multicolumn{3}{|c|}{n/a} & 4658.0 $\pm$ 0.9 & 5037.6 $\pm$ 1.1 & 0.925 \\
\hline
\end{tabular}
\label{tab::PerformanceRuns}
\end{table*}

Table \ref{tab::PerformanceRuns} provides multiple metrics of interest:
\begin{itemize}
    \item Run times of the current deposition kernel for GPU runs and for all macro-particle assignment-functions.
    \item Average time per total simulation step and their uncertainty.
    \item Calculated speedup with respect to the total simulation step duration when using EZ relative to Esirkepov's method.
    Speedup is the fraction of the average time per step required by Esirkepov's method in contrast to EZ.
    A  value greater than 1 indicates an improved performance with EZ.
\end{itemize}

Run time of the current deposition kernel is obtained by profiling with \texttt{ncu} and \texttt{rocprof} on NVIDIA V100 and AMD MI100, respectively.
Each data point of the average time per simulation step is an average of ten simulation runs executing one hundred steps.
The uncertainty of the mean value is determined from estimated standard deviation of sample data multiplied with a coverage factor of two to reach a level of confidence of approximately 95\,\% for the time per simulation step to lie in the given interval (assuming measurement results are normally distributed).

For all macro-particle assignment-functions, EZ decreases run time of the current deposition kernel as measured on GPUs, except for the highest order assignment-function PQS on AMD MI100.

With current deposition being just one part of the total simulation step, average run times of total steps provide a better measure for overall time savings in a particle-in-cell simulation than just kernel run times.
For this metric, EZ achieves a speedup on all three compute architectures.

For the first- and second-order macro-particle assignment-functions, CIC and TSC respectively, EZ consistently achieved speedup between 3.4\,\% and 13.7\,\%.
For the third-order macro-particle assignment function PQS, EZ's speedup varies across compute architectures.
For PQS on NVIDIA V100, 11.3\,\% speedup with EZ is the largest among all measured speedups on this compute architecture.
However, for PQS on AMD MI100, there is almost no performance difference.
Total simulation step run times differ by fewer than 1\,\%.
Only for PQS on AMD EPYC CPU, EZ did not achieve comparable or better performance compared to Esirkepov's method.

\subsubsection{Discussion}
On GPUs, the observed differences between EZ and optimized Esirkepov's method in kernel run times are predominantly in line with our expectations from the scaling of compute and memory complexity, $2\times(l+1)^2\times l$ for EZ and $(l+2)^2\times(l+1)$ for optimized Esirkepov's method, where we expect and observe the largest difference for CIC, a smaller difference for TSC and a small or no difference for PQS.
Only the much shorter run time with EZ for PQS on NVIDIA V100 is not in line with our expectation.

The profiles provide additional insight into the observed scaling of kernel run times.
On AMD MI100, Esirkepov's method performs more scalar Arithmetic Logic Unit (ALU) instructions and shared memory operations than EZ for all three assignment functions as reported in Table \ref{table:AMD-instr}.
\begin{table}[t]
\caption{
  Relative comparison of performance relevant profiling metrics between optimized Esirkepov's Method and EZ on AMD MI100.
  EZ is chosen as the baseline.}
  \centering
  \begin{tabular}{|c|c|c|c|}
  \hline
  Assignment & Scalar ALU Instr.\ & Shared Memory Ops.\ & Vector ALU Instr.\ \\
  function & \texttt{SALUInsts} & \texttt{LDSInsts} & \texttt{VALUInsts} \\
  \hline
  CIC & + 89\,\% & + 75\,\%  & + 29\,\% \\
  TSC & + 47\,\% & + 21\,\% & - 2\,\% \\
  PQS & + 10\,\%  & + 0.03\,\% & - 8\,\%  \\[1ex]
  \hline
  \end{tabular}
  \label{table:AMD-instr}
  \end{table}

For the slightly better performance of optimized Esirkepov's method with PQS on AMD MI100, the number of ALU calculations seems to make the critical difference due to the high arithmetic intensity for this assignment function at equal GPU occupancy of EZ and Esirkepov.
High arithmetic and low memory intensity is indicated by the \texttt{MemUnitBusy} metric telling that memory operations take only about 3\,\% of the total GPU time.
Equal GPU occupancy is indicated by equal usage of registers, as well as equal number of wavefronts of EZ and Esirkepov.
Summing vector and scalar instructions, Esirkepov's method performs about 6000 instructions (3\,\%) fewer than EZ, leading to fewer than 2\,\% kernel run time difference for PQS on AMD MI100.

On NVIDIA V100, a very similar scaling compared to AMD MI100 is observed in terms of shared memory instructions, arithmetic intensity, and occupancy.
Again the total number of executed instructions is the striking difference between Esirkepov's method and EZ.
Optimized Esirkepov's method performed more instructions for all three assignment functions (Table \ref{table:NVIDIA-instr}).
The profile also shows that the larger number of instructions for Esirkepov's method originates from more \texttt{branch instructions} compared to EZ.

\begin{table}[h!]
\caption{Relative comparison of performance relevant profiling metrics between optimized Esirkepov's Method and EZ on NVIDIA V100.
EZ is chosen as the baseline.}
\centering
\begin{tabular}{|c|c|c|}
\hline
Assignment & \texttt{executed} & \texttt{branch} \\
function & \texttt{instructions} & \texttt{instructions} \\
\hline
CIC & +88\,\% & +90\,\% \\
TSC & +96\,\% & +50\,\% \\
PQS & +116\,\%  & +18\,\% \\
\hline
\end{tabular}
\label{table:NVIDIA-instr}
\end{table}

As the implementation of the current deposition and the macro-particle assignment-functions are independent of compute architecture in PIConGPU, we argue that the observed performance difference for PQS, for which we expect comparable calculation and memory complexity, highly depends on compiler optimizations of the current deposition kernel and possibly of the different assignment-functions.
While the assignment-functions are mainly a series of \texttt{if}-conditions, with one condition fewer for EZ reflecting the on-support optimization, the current deposition consists of a nested triple \texttt{for}-loop and additional \texttt{if}-conditions providing room for a number of optimizations by the compiler such as unrolling or making use of predicates.

From the performance and profiling results of this exemplary real world simulation, we conclude that EZ provides, as expected, a substantial speedup for assignment-functions of order two and lower independent of compute architecture.
For assignment-functions of order three, EZ's performance depends on the details of the specific implementation, compute architecture and respective compiler implementations.

\section{Conclusions}
This work has presented a novel current decomposition method, EZ.
We demonstrated its generality and utility for different macro-particle assignment functions within the performance-portable particle-in-cell code PIConGPU.
We validated its charge conservation using standard deviation of the normalized Gauss' law remainder, showing that it has properties comparable to Esirkepov's method.

In addition, we verified the efficiency of EZ on several compute architectures and different accelerators by measuring performance, taking profiles, and comparing metrics to an optimized implementation of Esirkepov's method.

Performance measurements include current deposition kernel run time and average run time of a full particle-in-cell simulation step in a representative warm plasma simulation.

As expected from the scaling provided in the text, we observe a substantial speedup in these quantities with EZ on NVIDIA V100 and AMD MI100 GPUs, as well as AMD EPYC CPU for first (CIC) and second-order (TSC) macro-particle assignment-functions.
For the third-order macro-particle assignment-function PQS, where we expected the performance of EZ and Esirkepov's method to be equal, we found that speedup strongly depends on the choice of compute architecture, compilers, and optimizations.
For PQS, the highest order assignment function, the speedup of EZ over Esirkepov's method varied between lower (CPU), comparable (AMD MI100), and 11\,\% increased (NVIDIA V100) performance.
Notably, we used just one implementation for each current deposition method.
Due to usage of the performance portability library alpaka in PIConGPU, there is no need to implement hardware specific specializations of the method.

Beyond the direct performance improvements demonstrated within PIConGPU, every parallel implementation of the particle-in-cell method can benefit from EZ.
It ensures a uniform instruction sequence in the computation of the current density between all macro-particles, independent of the choice of macro-particle assignment-function.
This improves concurrency in the parallel computation of current density from many macro-particles.
More generally, it promotes the design of novel workload distribution algorithms for cooperative execution units running in parallel, such as warps on GPUs and SIMD instructions on CPUs.

EZ allows scientists to use high-performance compute systems with comparable or higher efficiency than Esirkepov's method without loss of precision for all but one of the considered cases.
This increase in efficiency enables a higher degree of sophistication in large-scale simulation campaigns modeling digital twins of lab experiments.
Such developments have many applications for future work, such as predicting the femtosecond and nanometer scale dynamics of laser-driven electron or proton accelerators with increased precision by exploiting the efficiency increase to run higher resolution simulations.
Likewise, the efficiency increase can be exploited to run more simulations in order to explore more of the possible parameter space of an experiment and identify interesting parameter regions to perform these experiments.

\section*{Acknowledgments}
This work was partly funded by the Center for Advanced Systems Understanding (CASUS) which is financed by Germany’s Federal Ministry of Education and Research (BMBF) and by the Saxon Ministry for Science, Culture and Tourism (SMWK) with tax funds on the basis of the budget approved by the Saxon State Parliament.

This research used resources of the Oak Ridge Leadership Computing Facility, which is a DOE Office of Science User Facility supported under Contract DE-AC05-00OR22725.


\bibliography{bib}

\pagebreak

\appendix

\input{supplement.tex}

\end{document}

%% file: supplement.tex

\section*{Supplemental Material}

Here we show that the expressions for the current density of a single macro-particle differ between the zigzag-line scheme and Esirkepov's method.

Starting with a macro-particle at position $(x_1, y_1, z_1)$ with assignment function CIC,
\[
  S_i(x_1) = \left\{\begin{array}{ll}1-| \xi|\,\text{, with } \xi=\tfrac{x_1}{\Delta x} - i & \text{for } 0\leqslant | \xi| < 1\\ 0 & \mathrm{Otherwise}\end{array}\right. \,,
\]
we recapitulate the calculation of the current density on the grid due to macro-particle movement to $(x_2, y_2, z_2)$.

\begin{figure}[tbp]
\centering
\includegraphics[width=0.75\textwidth]{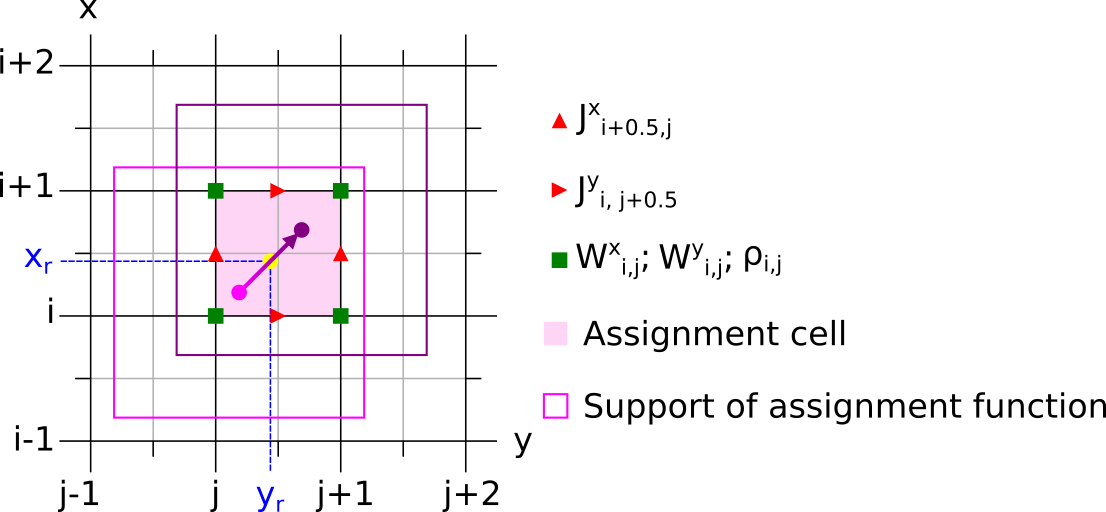}
\caption{Movement of a macro-particle with assignment function CIC on a 2D grid.
  The outline depicts the support of the macro-particle’s assignment function.
  During movement, the macro-particle stays within its assignment cell $[i, i + 1) \times [j, j + 1)]$.
  Grid nodes on which the x and y components of the current deposition vector $\mathbf W$ are non-zero are highlighted.
  The same holds true for the $x$ and $y$ components of the current density vector $\mathbf J$ on the grid edges.
  The coordinates of the splitting point in the zigzag-line scheme are denoted by $(x_r, y_r)$.
}
\label{fig::AssignmentSketch}
\end{figure}

It is assumed, that the macro-particle stays within the assignment cell after the movement.
That is, the grid index $i$ has a lower limit $i_\mathrm{min} = \mathrm{floor}(x_1/\Delta x) = \mathrm{floor}(x_2/\Delta x)$, which also marks the origin of the assignment cell.
The CIC assignment function thus becomes $S_i(x) = 1 + i - x$ for $i=i_\mathrm{min}, i_\mathrm{min} + 1$ and zero otherwise, similarly for the other axes.
See fig.\ \ref{fig::AssignmentSketch} for a 2D sketch of the grid with the macro-particle movement and non-vanishing values of current densities and the current distribution vector $\mathbf W$.

While the zigzag-line scheme splits the trajectory at $(x_1 + x_2)/2$ (similarly for the other axes) and calculates the total current density from two respective macro-particle movements, Esirkepov's method calculates the density from the total movement.
As is shown in the following, the methods provide different calculation rules.
\begin{enumerate}
\item Umeda, T., et al. ``A new charge conservation method in electromagnetic particle-in-cell simulations.'' \emph{Comp.\ Phys.\ Comm.} 156.1 (2003): 80, eq.\ (20)
  \begin{align*}
    J_x(i + \tfrac{1}{2}, j, k)& = \frac{q}{\Delta x \Delta y \Delta z}
        \left(\frac{x_r - x_1}{\Delta t}\right)
        \left( 1 + j - \frac{y_1 + y_r}{2\Delta y} \right)
        \left( 1 + k  - \frac{z_1+z_r}{2\Delta z} \right)\\
      &\quad + \frac{q}{\Delta x \Delta y \Delta z}\left(\frac{x_2 - x_r}{\Delta t}\right)
        \left( 1 + j - \frac{y_r + y_2}{2\Delta y} \right)
        \left( 1 + k - \frac{z_r + z_2}{2\Delta z} \right)\\
    & = \frac{q}{\Delta t \Delta y \Delta z}
        \left(\frac{x_2 - x_1}{2\Delta x}\right)
        \left( 1 + j - \frac{3y_1 + y_2}{4\Delta y} \right)
        \left( 1 + k - \frac{3z_1+z_2}{4\Delta z} \right)\\
      &\quad + \frac{q}{\Delta t \Delta y \Delta z}
        \left(\frac{x_2 - x_1}{2\Delta x}\right)
        \left( 1 + j - \frac{y_1 + 3y_2}{4\Delta y} \right)
        \left( 1 + k - \frac{z_1 + 3z_2}{4\Delta z} \right)\\
    & = \frac{q}{\Delta t \Delta y \Delta z}
        \frac{1}{2}
        \left(\frac{x_2 - x_1}{\Delta x}\right)\\
      &\quad \times
        \left[
          \left( 1 + j - \frac{3y_1 + y_2}{4\Delta y} \right)
            \left( 1 + k - \frac{3z_1+z_2}{4\Delta z} \right)
          + \left( 1 + j - \frac{y_1 + 3y_2}{4\Delta y} \right)
            \left( 1 + k - \frac{z_1 + 3z_2}{4\Delta z} \right)
        \right]
        \\
    & = \frac{q}{\Delta t \Delta y \Delta z}
        \frac{1}{2}
        \left(\frac{x_2 - x_1}{\Delta x}\right)\\
      &\quad \times
        \left[
          2(1+j)(1+k)
          - (1+k)\frac{y_1 + y_2}{\Delta y}
          - (1+j)\frac{z_1 + z_2}{\Delta z}
          \right. \\ &\qquad \left.
          + \frac{1}{16 \Delta y \Delta z}(10 y_1 z_1 + 6 y_1 z_2 + 6 y_2 z_1 + 10 y_2 z_2)
        \right]
\end{align*}
\item Esirkepov, T.\ Zh.\ ``Exact charge conservation scheme for particle-in-cell simulation with an arbitrary form-factor.'' \emph{Comp.\ Phys.\ Comm.} 135.2 (2001): 148, eq.\ (23)
  \begin{align*}
    J_x(i + \tfrac{1}{2}, j, k)& = -\frac{q}{\Delta t \Delta y \Delta z}\frac{(S_i(x_2) - S_i(x_1))}{6} \cdot \left\{
        S_k(z_2)[S_j(y_1) + 2S_j(y_2)] + [2S_j(y_1) + S_j(y_2)]S_k(z_1)
      \right\}\\
    & = \frac{q}{\Delta t \Delta y \Delta z} \frac{1}{6} \left( \frac{x_2 - x_1}{\Delta x} \right)\\
      & \quad \times \left[
          \left( 1 - \frac{z_2}{\Delta z} + k \right)
            \left( 1 - \frac{y_1 + 2y_2}{3\Delta y} + j \right) 3
          + 3 \left( 1 - \frac{2y_1 + y_2}{3\Delta y} + j \right)
            \left( 1 - \frac{z_1}{\Delta z} + k \right)
        \right]\\
    & = \frac{q}{\Delta t \Delta y \Delta z} \frac{1}{2} \left( \frac{x_2 - x_1}{\Delta x} \right)\\
      & \quad \times \left[
          \left( 1 - \frac{z_2}{\Delta z} + k \right)
            \left( 1 - \frac{y_1 + 2y_2}{3\Delta y} + j \right)
          + \left( 1 - \frac{2y_1 + y_2}{3\Delta y} + j \right)
            \left( 1 - \frac{z_1}{\Delta z} + k \right)
        \right]\\
    & = \frac{q}{\Delta t \Delta y \Delta z} \frac{1}{2} \left( \frac{x_2 - x_1}{\Delta x} \right)\\
      & \quad \times \left[
          2(1+j)(1+k)
          - (1+k)\frac{y_1 + y_2}{\Delta y}
          - (1+j)\frac{z_1 + z_2}{\Delta z}
          \right. \\ &\qquad \left.
          + \frac{1}{3 \Delta y \Delta z} (2 y_1 z_1 + y_1 z_2 + y_2 z_1 + 2 y_2 z_2)
        \right]
  \end{align*}
\end{enumerate}
These two calculation rules are only equal for 2D-like motions, where $z_1=z_2$ or $y_1=y_2$, but not for full 3D motions.

%% file: manuscript.bbl
\begin{thebibliography}{10}
\expandafter\ifx\csname url\endcsname\relax
  \def\url#1{\texttt{#1}}\fi
\expandafter\ifx\csname urlprefix\endcsname\relax\def\urlprefix{URL }\fi
\expandafter\ifx\csname href\endcsname\relax
  \def\href#1#2{#2} \def\path#1{#1}\fi

\bibitem{hockney1988}
R.~W. Hockney, J.~W. Eastwood, {Computer Simulation Using Particles (1st ed.)},
  {CRC Press}, 1988.
\newblock \href {http://dx.doi.org/10.1201/9780367806934}
  {\path{doi:10.1201/9780367806934}}.

\bibitem{birdsall1991}
C.~K. Birdsall, A.~B. Langdon, {Plasma Physics via Computer Simulation (1st
  ed.)}, CRC press, 1991.
\newblock \href {http://dx.doi.org/10.1201/9781315275048}
  {\path{doi:10.1201/9781315275048}}.

\bibitem{Fonseca2008}
R.~A. Fonseca, S.~F. Martins, L.~O. Silva, J.~W. Tonge, F.~S. Tsung, W.~B.
  Mori, {One-to-one direct modeling of experiments and astrophysical scenarios:
  pushing the envelope on kinetic plasma simulations}, Plasma Physics and
  Controlled Fusion 50~(12) (2008) 124034.
\newblock \href {http://arxiv.org/abs/0810.2460} {\path{arXiv:0810.2460}},
  \href {http://dx.doi.org/10.1088/0741-3335/50/12/124034}
  {\path{doi:10.1088/0741-3335/50/12/124034}}.

\bibitem{Burau2010}
H.~Burau, R.~Widera, W.~H{\"{o}}nig, G.~Juckeland, A.~Debus, T.~Kluge,
  U.~Schramm, T.~E. Cowan, R.~Sauerbrey, M.~Bussmann, {PIConGPU: A fully
  relativistic particle-in-cell code for a GPU cluster}, IEEE Transactions on
  Plasma Science 38~(10 PART 2) (2010) 2831--2839.
\newblock \href {http://dx.doi.org/10.1109/TPS.2010.2064310}
  {\path{doi:10.1109/TPS.2010.2064310}}.

\bibitem{Lehe2015a}
R.~Lehe, M.~Kirchen, I.~A. Andriyash, B.~B. Godfrey, J.~L. Vay, {A spectral,
  quasi-cylindrical and dispersion-free Particle-In-Cell algorithm}, Computer
  Physics Communications 203 (2016) 66--82.
\newblock \href {http://arxiv.org/abs/1507.04790} {\path{arXiv:1507.04790}},
  \href {http://dx.doi.org/10.1016/j.cpc.2016.02.007}
  {\path{doi:10.1016/j.cpc.2016.02.007}}.

\bibitem{Couperus2017}
J.~P. Couperus, R.~Pausch, A.~K{\"{o}}hler, O.~Zarini, J.~M. Kr{\"{a}}mer,
  M.~Garten, A.~Huebl, R.~Gebhardt, U.~Helbig, S.~Bock, K.~Zeil, A.~Debus,
  M.~Bussmann, U.~Schramm, A.~Irman,
  \href{http://www.nature.com/articles/s41467-017-00592-7}{{Demonstration of a
  beam loaded nanocoulomb-class laser wakefield accelerator}}, Nature
  Communications 8~(1) (2017) 487.
\newblock \href {http://dx.doi.org/10.1038/s41467-017-00592-7}
  {\path{doi:10.1038/s41467-017-00592-7}}.
\newline\urlprefix\url{http://www.nature.com/articles/s41467-017-00592-7}

\bibitem{Obst2018}
L.~Obst-Huebl, T.~Ziegler, F.-E. Brack, J.~Branco, M.~Bussmann, T.~E. Cowan,
  C.~B. Curry, F.~Fiuza, M.~Garten, M.~Gauthier, S.~G{\"{o}}de, S.~H. Glenzer,
  A.~Huebl, A.~Irman, J.~B. Kim, T.~Kluge, S.~D. Kraft, F.~Kroll,
  J.~Metzkes-Ng, R.~Pausch, I.~Prencipe, M.~Rehwald, C.~Roedel, H.-P.
  Schlenvoigt, U.~Schramm, K.~Zeil,
  \href{http://www.nature.com/articles/s41467-018-07756-z}{{All-optical
  structuring of laser-driven proton beam profiles}}, Nature Communications
  9~(1) (2018) 5292.
\newblock \href {http://dx.doi.org/10.1038/s41467-018-07756-z}
  {\path{doi:10.1038/s41467-018-07756-z}}.
\newline\urlprefix\url{http://www.nature.com/articles/s41467-018-07756-z}

\bibitem{Derouillat2018}
J.~Derouillat, A.~Beck, F.~P{\'{e}}rez, T.~Vinci, M.~Chiaramello, A.~Grassi,
  M.~Fl{\'{e}}, G.~Bouchard, I.~Plotnikov, N.~Aunai, J.~Dargent, C.~Riconda,
  M.~Grech,
  \href{https://linkinghub.elsevier.com/retrieve/pii/S0010465517303314}{{Smilei
  : A collaborative, open-source, multi-purpose particle-in-cell code for
  plasma simulation}}, Computer Physics Communications 222 (2018) 351--373.
\newblock \href {http://arxiv.org/abs/1702.05128} {\path{arXiv:1702.05128}},
  \href {http://dx.doi.org/10.1016/j.cpc.2017.09.024}
  {\path{doi:10.1016/j.cpc.2017.09.024}}.
\newline\urlprefix\url{https://linkinghub.elsevier.com/retrieve/pii/S0010465517303314}

\bibitem{Debus2017a}
A.~Debus, R.~Pausch, A.~Huebl, K.~Steiniger, R.~Widera, T.~E.~T. Cowan,
  U.~Schramm, M.~Bussmann,
  \href{https://link.aps.org/doi/10.1103/PhysRevX.9.031044}{{Circumventing the
  Dephasing and Depletion Limits of Laser-Wakefield Acceleration}}, Physical
  Review X 9~(3) (2019) 031044.
\newblock \href {http://dx.doi.org/10.1103/PhysRevX.9.031044}
  {\path{doi:10.1103/PhysRevX.9.031044}}.
\newline\urlprefix\url{https://link.aps.org/doi/10.1103/PhysRevX.9.031044}

\bibitem{Raj2020}
G.~Raj, O.~Kononenko, M.~F. Gilljohann, A.~Doche, X.~Davoine, C.~Caizergues,
  Y.-Y. Chang, J.~P. {Couperus Cabadağ}, A.~Debus, H.~Ding, M.~F{\"{o}}rster,
  J.-P. Goddet, T.~Heinemann, T.~Kluge, T.~Kurz, R.~Pausch, P.~Rousseau,
  P.~{San Miguel Claveria}, S.~Sch{\"{o}}bel, A.~Siciak, K.~Steiniger,
  A.~Tafzi, S.~Yu, B.~Hidding, A.~{Martinez de la Ossa}, A.~Irman, S.~Karsch,
  A.~D{\"{o}}pp, U.~Schramm, L.~Gremillet, S.~Corde,
  \href{https://link.aps.org/doi/10.1103/PhysRevResearch.2.023123}{{Probing
  ultrafast magnetic-field generation by current filamentation instability in
  femtosecond relativistic laser-matter interactions}}, Physical Review
  Research 2~(2) (2020) 023123.
\newblock \href {http://dx.doi.org/10.1103/PhysRevResearch.2.023123}
  {\path{doi:10.1103/PhysRevResearch.2.023123}}.
\newline\urlprefix\url{https://link.aps.org/doi/10.1103/PhysRevResearch.2.023123}

\bibitem{Kurz2020}
T.~Kurz, T.~Heinemann, M.~F. Gilljohann, Y.~Y. Chang, J.~{Couperus Cabadağ},
  A.~Debus, O.~Kononenko, R.~Pausch, S.~Sch{\"{o}}bel, R.~W. Assmann,
  M.~Bussmann, H.~Ding, J.~G{\"{o}}tzfried, A.~K{\"{o}}hler, G.~Raj,
  S.~Schindler, K.~Steiniger, O.~Zarini, S.~Corde, A.~D{\"{o}}pp, B.~Hidding,
  S.~Karsch, U.~Schramm, A.~{Martinez De La Ossa}, A.~Irman, {Demonstration of
  a compact plasma accelerator powered by laser-accelerated electron beams},
  Nature Communications 12~(1).
\newblock \href {http://arxiv.org/abs/1909.06676} {\path{arXiv:1909.06676}},
  \href {http://dx.doi.org/10.1038/s41467-021-23000-7}
  {\path{doi:10.1038/s41467-021-23000-7}}.

\bibitem{CouperusCabadag2021}
J.~P. {Couperus Cabadağ}, R.~Pausch, S.~Sch{\"{o}}bel, M.~Bussmann, Y.-Y.
  Chang, S.~Corde, A.~Debus, H.~Ding, A.~D{\"{o}}pp, F.~M. Foerster,
  M.~Gilljohann, F.~Haberstroh, T.~Heinemann, B.~Hidding, S.~Karsch,
  A.~Koehler, O.~Kononenko, A.~Knetsch, T.~Kurz, A.~{Martinez de la Ossa},
  A.~Nutter, G.~Raj, K.~Steiniger, U.~Schramm, P.~Ufer, A.~Irman,
  \href{www.hzdr.de/fwt
  https://link.aps.org/doi/10.1103/PhysRevResearch.3.L042005}{{Gas-dynamic
  density downramp injection in a beam-driven plasma wakefield accelerator}},
  Physical Review Research 3~(4) (2021) L042005.
\newblock \href {http://dx.doi.org/10.1103/PhysRevResearch.3.L042005}
  {\path{doi:10.1103/PhysRevResearch.3.L042005}}.
\newline\urlprefix\url{www.hzdr.de/fwt
  https://link.aps.org/doi/10.1103/PhysRevResearch.3.L042005}

\bibitem{Haugbolle2013}
T.~Haugb{\o}lle, J.~T. Frederiksen, {\AA}.~Nordlund,
  \href{http://aip.scitation.org/doi/10.1063/1.4811384}{{Photon-Plasma: A
  modern high-order particle-in-cell code}}, Physics of Plasmas 20~(6) (2013)
  062904.
\newblock \href {http://arxiv.org/abs/arXiv:1211.4575v3}
  {\path{arXiv:arXiv:1211.4575v3}}, \href {http://dx.doi.org/10.1063/1.4811384}
  {\path{doi:10.1063/1.4811384}}.
\newline\urlprefix\url{http://aip.scitation.org/doi/10.1063/1.4811384}

\bibitem{Pausch2014a}
R.~Pausch, H.~Burau, M.~Bussmann, J.~Couperus, T.~E. Cowan, A.~Debus, A.~Huebl,
  A.~Irman, A.~K{\"{o}}hler, U.~Schramm, K.~Steiniger, R.~Widera,
  \href{http://accelconf.web.cern.ch/AccelConf/IPAC2014/papers/mopri069.pdf}{{Computing
  angularly-resolved far field emission spectra in particle-in-cell codes using
  GPUs}}, in: Proceeding of IPAC2014, Vol. MOPRI069, 2014, pp. 761--764.
\newblock \href {http://dx.doi.org/10.18429/JACoW-IPAC2014-MOPRI069}
  {\path{doi:10.18429/JACoW-IPAC2014-MOPRI069}}.
\newline\urlprefix\url{http://accelconf.web.cern.ch/AccelConf/IPAC2014/papers/mopri069.pdf}

\bibitem{Pukhov1999}
A.~Pukhov,
  \href{https://www.cambridge.org/core/product/identifier/S0022377899007515/type/journal_article}{{Three-dimensional
  electromagnetic relativistic particle-in-cell code VLPL (Virtual Laser Plasma
  Lab)}}, Journal of Plasma Physics 61~(3) (1999) 425--433.
\newblock \href {http://dx.doi.org/10.1017/S0022377899007515}
  {\path{doi:10.1017/S0022377899007515}}.
\newline\urlprefix\url{https://www.cambridge.org/core/product/identifier/S0022377899007515/type/journal_article}

\bibitem{Nishikawa2009}
K.-I. Nishikawa, M.~Medvedev, B.~Zhang, P.~Hardee, J.~Niemiec, {\AA}.~Nordlund,
  J.~Frederiksen, Y.~Mizuno, H.~Sol, G.~J. Fishman, C.~Meegan, C.~Kouveliotou,
  N.~Gehrels,
  \href{http://aip.scitation.org/doi/abs/10.1063/1.3155889}{{Radiation from
  relativistic jets in turbulent magnetic fields}}, in: AIP Conference
  Proceedings, Vol. 1133, AIP, 2009, pp. 235--237.
\newblock \href {http://dx.doi.org/10.1063/1.3155889}
  {\path{doi:10.1063/1.3155889}}.
\newline\urlprefix\url{http://aip.scitation.org/doi/abs/10.1063/1.3155889}

\bibitem{Bussmann2013}
M.~Bussmann, H.~Burau, T.~E. Cowan, A.~Debus, A.~Huebl, G.~Juckeland, T.~Kluge,
  W.~E. Nagel, R.~Pausch, F.~Schmitt, U.~Schramm, J.~Schuchart, R.~Widera,
  \href{http://dl.acm.org/citation.cfm?doid=2503210.2504564}{{Radiative
  signatures of the relativistic Kelvin-Helmholtz instability}}, in: SC '13
  Proceedings of the International Conference for High Performance Computing,
  Networking, Storage and Analysis, 2013, pp. 5--1 -- 5--12.
\newblock \href {http://dx.doi.org/10.1145/2503210.2504564}
  {\path{doi:10.1145/2503210.2504564}}.
\newline\urlprefix\url{http://dl.acm.org/citation.cfm?doid=2503210.2504564}

\bibitem{Grismayer2013a}
T.~Grismayer, E.~P. Alves, R.~a. Fonseca, L.~O. Silva,
  \href{https://link.aps.org/doi/10.1103/PhysRevLett.111.015005}{{dc-Magnetic-Field
  Generation in Unmagnetized Shear Flows}}, Physical Review Letters 111~(1)
  (2013) 015005.
\newblock \href {http://dx.doi.org/10.1103/PhysRevLett.111.015005}
  {\path{doi:10.1103/PhysRevLett.111.015005}}.
\newline\urlprefix\url{https://link.aps.org/doi/10.1103/PhysRevLett.111.015005}

\bibitem{PauschPRE2017}
R.~Pausch, M.~Bussmann, A.~Huebl, U.~Schramm, K.~Steiniger, R.~Widera,
  A.~Debus,
  \href{http://link.aps.org/doi/10.1103/PhysRevE.96.013316}{{Identifying the
  linear phase of the relativistic Kelvin-Helmholtz instability and measuring
  its growth rate via radiation}}, Physical Review E 96~(1) (2017) 013316.
\newblock \href {http://dx.doi.org/10.1103/PhysRevE.96.013316}
  {\path{doi:10.1103/PhysRevE.96.013316}}.
\newline\urlprefix\url{http://link.aps.org/doi/10.1103/PhysRevE.96.013316}

\bibitem{Hilz2018}
P.~Hilz, T.~M. Ostermayr, A.~Huebl, V.~Bagnoud, B.~Borm, M.~Bussmann,
  M.~Gallei, J.~Gebhard, D.~Haffa, J.~Hartmann, T.~Kluge, F.~H. Lindner,
  P.~Neumayr, C.~G. Schaefer, U.~Schramm, P.~G. Thirolf, T.~F. R{\"{o}}sch,
  F.~Wagner, B.~Zielbauer, J.~Schreiber, {Isolated proton bunch acceleration by
  a petawatt laser pulse}, Nature Communications 9~(1) (2018) 423.
\newblock \href {http://dx.doi.org/10.1038/s41467-017-02663-1}
  {\path{doi:10.1038/s41467-017-02663-1}}.

\bibitem{Myers2021}
A.~Myers, A.~Almgren, L.~Amorim, J.~Bell, L.~Fedeli, L.~Ge, K.~Gott, D.~Grote,
  M.~Hogan, A.~Huebl, R.~Jambunathan, R.~Lehe, C.~Ng, M.~Rowan, O.~Shapoval,
  M.~Th{\'{e}}venet, J.-L. Vay, H.~Vincenti, E.~Yang, N.~Za{\"{i}}m, W.~Zhang,
  Y.~Zhao, E.~Zoni,
  \href{https://www.sciencedirect.com/science/article/pii/S0167819121000818}{{Porting
  WarpX to GPU-accelerated platforms}}, Parallel Computing 108 (2021) 102833.
\newblock \href {http://dx.doi.org/10.1016/j.parco.2021.102833}
  {\path{doi:10.1016/j.parco.2021.102833}}.
\newline\urlprefix\url{https://www.sciencedirect.com/science/article/pii/S0167819121000818}

\bibitem{leinhauser2022}
M.~Leinhauser, R.~Widera, S.~Bastrakov, A.~Debus, M.~Bussmann,
  S.~Chandrasekaran, \href{https://doi.org/10.1145/3505285}{Metrics and design
  of an instruction roofline model for amd gpus}, ACM Trans. Parallel Comput.
  9~(1).
\newblock \href {http://dx.doi.org/10.1145/3505285}
  {\path{doi:10.1145/3505285}}.
\newline\urlprefix\url{https://doi.org/10.1145/3505285}

\bibitem{eastwood1991}
J.~W. Eastwood,
  \href{https://www.sciencedirect.com/science/article/pii/001046559190036K}{The
  virtual particle electromagnetic particle-mesh method}, Computer Physics
  Communications 64~(2) (1991) 252--266.
\newblock \href {http://dx.doi.org/10.1016/0010-4655(91)90036-K}
  {\path{doi:10.1016/0010-4655(91)90036-K}}.
\newline\urlprefix\url{https://www.sciencedirect.com/science/article/pii/001046559190036K}

\bibitem{villasenor1992}
J.~Villasenor, O.~Buneman,
  \href{https://www.sciencedirect.com/science/article/pii/001046559290169Y}{Rigorous
  charge conservation for local electromagnetic field solvers}, Computer
  Physics Communications 69~(2) (1992) 306--316.
\newblock \href {http://dx.doi.org/10.1016/0010-4655(92)90169-Y}
  {\path{doi:10.1016/0010-4655(92)90169-Y}}.
\newline\urlprefix\url{https://www.sciencedirect.com/science/article/pii/001046559290169Y}

\bibitem{esirkepov2001}
T.~Esirkepov,
  \href{https://www.sciencedirect.com/science/article/pii/S0010465500002289}{{Exact
  charge conservation scheme for Particle-in-Cell simulation with an arbitrary
  form-factor}}, Computer Physics Communications 135~(2) (2001) 144--153.
\newblock \href {http://dx.doi.org/10.1016/S0010-4655(00)00228-9}
  {\path{doi:10.1016/S0010-4655(00)00228-9}}.
\newline\urlprefix\url{https://www.sciencedirect.com/science/article/pii/S0010465500002289}

\bibitem{umeda2003}
T.~Umeda, Y.~Omura, T.~Tominaga, H.~Matsumoto,
  \href{https://www.sciencedirect.com/science/article/pii/S0010465503004375}{A
  new charge conservation method in electromagnetic particle-in-cell
  simulations}, Computer Physics Communications 156~(1) (2003) 73--85.
\newblock \href {http://dx.doi.org/10.1016/S0010-4655(03)00437-5}
  {\path{doi:10.1016/S0010-4655(03)00437-5}}.
\newline\urlprefix\url{https://www.sciencedirect.com/science/article/pii/S0010465503004375}

\bibitem{praceWWW}
{Partnership For Advanced Computing in Europe ("PRACE")}, {PRACE 15th Call
  continues to award outstanding research in HPC},
  \url{https://prace-ri.eu/pr-prace-project-access-call15/}, [Online; accessed
  30-March-2022] (2017).

\bibitem{yu2013}
J.~Yu, X.~Jin, W.~Zhou, B.~Li, Y.~Gu, {High-Order Interpolation Algorithms for
  Charge Conservation in Particle-in-Cell Simulations}, Communications in
  Computational Physics 13~(4) (2013) 1134–1150.
\newblock \href {http://dx.doi.org/10.4208/cicp.290811.050412a}
  {\path{doi:10.4208/cicp.290811.050412a}}.

\bibitem{picongpuWWW}
{PIConGPU: Particle-in-Cell Simulations for the Exascale Era},
  \url{https://github.com/ComputationalRadiationPhysics/picongpu}, [Online;
  latest release picongpu-0.6.0 at 21-December-2021] (2021).

\bibitem{zenker2016}
E.~Zenker, B.~Worpitz, R.~Widera, A.~Huebl, G.~Juckeland, A.~Kn{\"{u}}pfer,
  W.~E. Nagel, M.~Bussmann, \href{http://arxiv.org/abs/1602.08477}{Alpaka - an
  abstraction library for parallel kernel acceleration}, IEEE Computer Society,
  2016.
\newblock \href {http://arxiv.org/abs/1602.08477} {\path{arXiv:1602.08477}}.
\newline\urlprefix\url{http://arxiv.org/abs/1602.08477}

\bibitem{matthes2017}
A.~{Matthes}, R.~{Widera}, E.~{Zenker}, B.~{Worpitz}, A.~{Huebl},
  M.~{Bussmann}, \href{https://arxiv.org/abs/1706.10086}{Tuning and
  optimization for a variety of many-core architectures without changing a
  single line of implementation code using the alpaka library}, 2017.
\newblock \href {http://arxiv.org/abs/1706.10086} {\path{arXiv:1706.10086}}.
\newline\urlprefix\url{https://arxiv.org/abs/1706.10086}

\bibitem{serialization2023}
{NVIDIA Corporation \& affiliates}, {CUDA C++ Best Practices Guide v12.1},
  \url{https://docs.nvidia.com/cuda/archive/12.0.1/cuda-c-best-practices-guide/index.html#branching-and-divergence},
  [Online; accessed 24-March-2023] (2022).

\bibitem{yee1966}
K.~Yee, {Numerical solution of initial boundary value problems involving
  Maxwell's equations in isotropic media}, IEEE Transactions on Antennas and
  Propagation 14~(3) (1966) 302--307.
\newblock \href {http://dx.doi.org/10.1109/TAP.1966.1138693}
  {\path{doi:10.1109/TAP.1966.1138693}}.

\bibitem{godfrey2013}
B.~B. Godfrey, J.-L. Vay,
  \href{https://www.sciencedirect.com/science/article/pii/S0021999113002556}{{Numerical
  stability of relativistic beam multidimensional PIC simulations employing the
  Esirkepov algorithm}}, Journal of Computational Physics 248 (2013) 33--46.
\newblock \href {http://dx.doi.org/https://doi.org/10.1016/j.jcp.2013.04.006}
  {\path{doi:https://doi.org/10.1016/j.jcp.2013.04.006}}.
\newline\urlprefix\url{https://www.sciencedirect.com/science/article/pii/S0021999113002556}

\bibitem{meyers2015}
M.~Meyers, C.-K. Huang, Y.~Zeng, S.~Yi, B.~Albright,
  \href{https://www.sciencedirect.com/science/article/pii/S002199911500371X}{{On
  the numerical dispersion of electromagnetic particle-in-cell code: Finite
  grid instability}}, Journal of Computational Physics 297 (2015) 565--583.
\newblock \href {http://dx.doi.org/https://doi.org/10.1016/j.jcp.2015.05.037}
  {\path{doi:https://doi.org/10.1016/j.jcp.2015.05.037}}.
\newline\urlprefix\url{https://www.sciencedirect.com/science/article/pii/S002199911500371X}

\bibitem{xiao2019}
J.~Xiao, H.~Qin, Y.~Shi, J.~Liu, R.~Zhang,
  \href{https://www.sciencedirect.com/science/article/pii/S0375960118312106}{{A
  lattice Maxwell system with discrete space–time symmetry and local
  energy–momentum conservation}}, Physics Letters A 383~(9) (2019) 808--812.
\newblock \href
  {http://dx.doi.org/https://doi.org/10.1016/j.physleta.2018.12.010}
  {\path{doi:https://doi.org/10.1016/j.physleta.2018.12.010}}.
\newline\urlprefix\url{https://www.sciencedirect.com/science/article/pii/S0375960118312106}

\bibitem{occupancy2023}
{NVIDIA Corporation \& affiliates}, {CUDA C++ Best Practices Guide v12.1},
  \url{https://docs.nvidia.com/cuda/archive/12.0.1/cuda-c-best-practices-guide/index.html#execution-configuration-optimizations},
  [Online; accessed 24-March-2023] (2023).

\bibitem{widera2022}
K.~Steiniger, R.~Widera, J.~Young, {EZ publication: source code, profiling,
  analysis and simulation data} (2022).
\newblock \href {http://dx.doi.org/10.14278/rodare.1511}
  {\path{doi:10.14278/rodare.1511}}.

\bibitem{zyla2020}
P.~Zyla, et~al., {Review of Particle Physics}, PTEP 2020~(8) (2020) 083C01,
  {Statistics Review; sec.\ 40.2}.
\newblock \href {http://dx.doi.org/10.1093/ptep/ptaa104}
  {\path{doi:10.1093/ptep/ptaa104}}.

\bibitem{summitWWW}
{Oakridge Leadership Computing Facility}, {Summit User Guide},
  \url{https://docs.olcf.ornl.gov/systems/summit_user_guide.html#summit-nodes},
  [Online; accessed 17-December-2021] (2021).

\bibitem{spockWWW}
{Oakridge Leadership Computing Facility}, {Spock Quick-Start Guide},
  \url{https://docs.olcf.ornl.gov/systems/spock_quick_start_guide.html#spock-compute-nodes},
  [Online; accessed 17-December-2021] (2021).

\end{thebibliography}
